\documentclass[preprint, 11pt]{elsarticle}
\usepackage{fullpage} 
\usepackage{graphicx}   
\usepackage{comment} 
\usepackage{amssymb,amsmath, mathrsfs,mathtools}
\usepackage{caption}
\usepackage{setspace}
\usepackage{multirow}
\usepackage{array}
\usepackage{xurl}
\usepackage{color}
\usepackage{longtable}
\usepackage{multirow}
\biboptions{sort&compress}
\usepackage[colorlinks,allcolors=blue]{hyperref}

\begin{document}
\begin{frontmatter}

\title{Mechanistic Modeling and Analysis of Thermal Radiation in Conventional, Microwave-assisted, and Hybrid Freeze Drying for Biopharmaceutical Manufacturing}

\author[MechE,ChemE,CCSE]{Prakitr Srisuma}
\author[MechE,CCSE]{George Barbastathis}
\author[ChemE,CCSE]{Richard D. Braatz\corref{coauthor}}
\cortext[coauthor]{Corresponding author}
\ead{braatz@mit.edu}

\address[MechE]{Department of Mechanical Engineering, Massachusetts Institute of Technology, Cambridge, MA 02139}
\address[ChemE]{Department of Chemical Engineering, Massachusetts Institute of Technology, Cambridge, MA 02139}
\address[CCSE]{Center for Computational Science and Engineering, Massachusetts Institute of Technology, Cambridge, MA 02139}


\begin{abstract}
In freeze drying, thermal radiation has a significant effect on the drying process of vials located near the corner and edge of the trays, resulting in non-uniformity of the products. Understanding and being able to predict the impact of thermal radiation are therefore critical to accurate determination of the drying process endpoint given the variation in heat transfer of each vial. This article presents a new mechanistic model that describes complex thermal radiation during primary drying in conventional, microwave-assisted, and hybrid freeze drying. Modeling of thermal radiation employs the diffuse gray surface model and radiation network approach, which systematically and accurately incorporates simultaneous radiation exchange between every surface including the chamber wall and vials, allowing the framework to be seamlessly applied for analyzing various freeze-dryer designs. Model validation with data from the literature shows accurate prediction of the drying times for all vials, including inner, edge, and corner vials. The validated model is demonstrated for thermal radiation analysis and parametric studies to guide the design and optimization of freeze dryers. 
\end{abstract}

\begin{keyword}
Lyophilization \sep Freeze drying \sep Thermal radiation \sep Diffuse gray surface \sep Radiation network \sep Monte Carlo simulation 
\end{keyword}

\end{frontmatter}

\section{Introduction} \label{sec:Intro}
Freeze drying, also known as lyophilization, is a key process used in the formulation of biotherapeutics. The process comprises three stages: freezing, primary drying, and secondary drying. During the freezing stage, the liquid solvent (typically water) is frozen at low temperature \cite{Fissore_2018_Review, Bano_2020_optimize}. The free water transforms into ice crystals, while the bound water remains in a non-crystalline state bound to the product molecules \cite{Fissore_2018_Review}. In primary drying, the frozen product and solvent are subjected to low pressure and temperature, causing the ice crystals to undergo sublimation \cite{Pisano_2010_control}. The subsequent stage, secondary drying, involves further heating of the product to higher temperature to remove most of the remaining bound water through desorption \cite{Veraldi_2008_Parameters}. In comparison to other drying techniques, freeze drying is performed at lower temperatures, making it particularly suitable for preserving the quality and structure of heat-sensitive materials, such as biopharmaceutical products \cite{Barresi_2009_Monitoring}. Recently, freeze drying has been shown to provide long-term stability for mRNA vaccines, which enables these vaccines to be delivered in countries that lack a cold supply chain \cite{Muramatsu_2022_mRNA,Meulewaeter_2023_mRNA}.

Conventional freeze drying (CFD) provides heat to the frozen product by means of a heating shelf positioned beneath the drying chamber or vial, that is, by heat conduction \cite{Pisano_2010_control,Fissore_2018_Review}. Microwave-assisted freeze drying (MFD) employs microwave irradiation to heat the product to reduce the drying time, and hybrid freeze drying (HFD) combines the heating techniques of both CFD and MFD  \cite{Gitter_2018_Experiment,Gitter_2019_Parameters,Bhambhani_2021_MVD,Richard_2021_MainModel}. Primary drying, known for its lengthy duration, potential hazards, and high costs, has become the primary focus for optimization efforts \cite{Veraldi_2008_Parameters,Pisano_2010_control}. Mechanistic models have been developed for these purposes, e.g., see examples and discussions by \cite{Litchfield_1979_Model,Mascarenhas_1997_FEMmodel,Pikal_2005_Model,Hottot_2006_Parameters,Veraldi_2008_Parameters,Nastaj_2009_MFD,Pisano_2010_control,Chen_2015_FEMmodel,Fissore_2015_HTCestimation,Scutella_2017_3Dmodel,Bano_2020_optimize,Wang_2020_MircowaveModel,Richard_2021_MainModel,Srisuma_2023_AnalyticalLyo}.

Besides heat conduction and microwave irradiation, another mode of heat transfer in primary drying is thermal radiation, which is due to the difference in temperature of the chamber wall and vials \cite{Pisano_2011_HeatTransferInLyo}. Thermal radiation usually has a significant impact on the outermost vials, whereas its effect is negligible for inner vials as these vials are shielded from the outer vials \cite{Veraldi_2008_Parameters,Fissore_2018_Review}. The additional heat from thermal radiation generally accelerates the drying process of the outer vials, leading to variation in the drying times \cite{Pikal_2016_Convection}. {\color{black}  In general, understanding and being able to predict the impact of thermal radiation are critical to accurate determination of the product temperature and drying process endpoint for each vial, especially located near the corner and edge of the trays. This insight can also be vital for the optimization of freeze-dryer design.} Although extensive experimental investigations have explored the effects of thermal radiation in freeze drying \cite{Rambhatla_2003_RadiationShield,Gan_2005_WallTemp,Veraldi_2008_Parameters,Pikal_2016_Convection,Bano_2020_optimize,Ehlers_2021_WallTemp}, the literature on mechanistic modeling for analyzing such effects is more limited. Published models for thermal radiation during primary drying assume radiation exchange exists between each vial and the chamber wall independently, and so approximate the radiative heat transfer with a simple function of the fourth power of the absolute temperature \cite{Sheehan_1998_Modeling,Gan_2005_WallTemp,Veraldi_2008_Parameters,Pikal_2016_Convection,Bano_2020_optimize} or in the form of Newton's law of cooling with the effective heat transfer coefficient \cite{Hottot_2006_Parameters,Pisano_2011_HeatTransferInLyo,Chen_2015_FEMmodel,Fissore_2015_HTCestimation}. In practice, thermal radiation exchange exists not only between the vial and chamber wall but also between multiple vials, where all of these processes occur simultaneously. Freeze drying of pharmaceutical and biopharmaceutical products usually entails a large number of vials \cite{Veraldi_2008_Parameters,Bano_2020_optimize,Bhambhani_2021_MVD}, which strengthens the effect of radiation exchange between vials. To our knowledge, an accurate model of this complicated phenomenon is not available in the literature.

This article presents a novel mechanistic model that accounts for complex thermal radiation exchange in primary drying for all types of freeze drying, including CFD, MFD, and HFD. Our framework relies on the diffuse gray surface model and radiation network approach, which systematically and accurately incorporates simultaneous radiation exchange between every surface including the chamber wall and vials. The model is validated with simulation and experimental studies from the literature. Applications of the model are demonstrated for analysis and parametric studies of thermal radiation in a freeze dryer.

This article is organized as follows. Section \ref{sec:Model} describes the mechanistic model for simulating primary drying without thermal radiation. Section \ref{sec:ThermalRadiation} derives the diffuse gray surface model and radiation network approach for modeling of thermal radiation. Section \ref{sec:ModelValiation} validates the model with data from the literature. Finally, Section \ref{sec:Analysis} employs the model for in-depth analysis of thermal radiation in primary drying.


\section{Mechanistic Model for Freeze Drying}  \label{sec:Model}
{\color{black} The mechanistic model used in this work is based primarily on the simplified model discussed in \cite{Srisuma_2023_AnalyticalLyo} with slight modifications; we refer to the aforementioned work for detailed derivation, solution methods, and simulations. This section firstly discusses some important findings in the literature and then summarizes the model to be used as a basis for extension to thermal radiation modeling in the next section.

The key phenomenon in primary drying is sublimation, and there are many modeling strategies proposed for this process in the literature. Sublimation is a simultaneous heat and mass transfer process \cite{Mills_1995_HeatTransfer}, and so mass and heat transfer are modeled together in some literature \cite{Litchfield_1979_Model,Sheehan_1998_Modeling,Veraldi_2008_Parameters,Gitter_2018_Experiment}. However, it has been observed in both simulations and experiments that, in primary drying, the product temperature increases significantly at the beginning and then becomes approximately constant after reaching some certain threshold \cite{Hottot_2006_Parameters,Veraldi_2008_Parameters,Gitter_2018_Experiment}. As a result, instead of modeling simultaneous heat and mass transfer, many models, including ours, approximate that sublimation is controlled by heat transfer only; i.e., the sublimation rate is directly controlled by the amount of heat input \cite{Dryer_1969_HeatTransferControlled,Jafar_2003_HeatTransferLimit,Hottot_2006_Parameters,Richard_2021_MainModel}, and the model prediction is proven to be sufficiently accurate. This simplification reduces the number of equations and parameters, simplifying calculation and real-time implementation of the model. In addition, the fact that microwave irradiation can reduce the drying time by about 
 80\% \cite{Gitter_2018_Experiment,Gitter_2019_Parameters} suggests that the process is primarily controlled by heat transfer.

Our model is formulated in the rectangular coordinate system with one spatial dimension ($x$) and time ($t$), which consists of two parts denoted as the (1) heating stage and (2) sublimation stage (Fig.\ \ref{fig:Model_Schematic}). The heating stage assumes no phase change in the system, and so the supplied heat increases the temperature of the frozen region. The sublimation stage describes the evolution of the sublimating interface, with the supplied heat assumed to be used for sublimation only. The heating stage exists at $0\leq t<t_m$, whereas the sublimation stage exists at $ t\geq t_m$, where $t_m$ is the time when sublimation starts. This two-stage model corresponds to the observation discussed in the previous paragraph, with the temperature threshold denoted by $T_m$, the sublimation (i.e, equilibrium) temperature represented by the solid-vapor line in the phase diagram.}

During the heating stage, the energy balance for the frozen region is
\begin{equation} \label{eq:governing_original}
    \rho C_P\dfrac{\partial T}{\partial t} = k\dfrac{\partial^2 T}{\partial x^2} + H_{v1}, \qquad  0 < x < L, \qquad 0<t<t_m,
\end{equation}
where $T(x,t)$ is the temperature, $H_{v1}$ is the microwave irradiation that affects the frozen material during the heating stage, $\rho$ is the density, $k$ is the thermal conductivity, $C_P$ is the heat capacity, and {\color{black} $L$ is the height of the frozen material.} 

The bottom surface of the frozen region is heated by the bottom shelf, following Newton's law of cooling
\begin{equation}\label{eq:boundary_bottom}
    -k\dfrac{\partial T}{\partial x}(L,t) = h(T(L,t)-T_b(t)), \qquad 0<t<t_m,
\end{equation}
where $h$ is the heat transfer coefficient at the bottom and $T_b(t)$ is the bottom shelf temperature. The heat transfer coefficient can be estimated from either correlations or experimental data. {\color{black} This heat transfer coefficient is usually treated as an effective heat transfer coefficient that accounts for three heat transfer mechanisms between the heating shelf and the bottom surface of the vial: (1) heat conduction at the point of contact, (2) convection from the gas phase, and (3) thermal radiation  \cite{Pikal_1984_HTCexperiment,Pikal_2005_Model,Veraldi_2008_Parameters,Pisano_2010_control}.} {\color{black} The shelf temperature is assumed to increase linearly as a function of time, 
\begin{equation}\label{eq:shelf_temp}
    T_{b}(t) = rt + T_{b0},
\end{equation}
where $T_{b0}$ is the initial shelf temperature and $r$ is the temperature ramp-up rate. After reaching the maximum temperature $T_{b,\textrm{max}}$, the shelf temperature is kept constant at that value. This linear temperature ramp-up strategy is relatively common in the literature, including all the case studies presented in this work \cite{Veraldi_2008_Parameters,Gitter_2018_Experiment,Gitter_2019_Parameters}. 

{\color{black} Past simulation and experimental studies have reported \cite{Pikal_2005_Model,Veraldi_2008_Parameters,Richard_2021_MainModel} that heat transfer is much weaker at the top surface than the bottom surface, in which case the boundary condition can be approximated as
\begin{equation} \label{eq:boundary_top}
    k\dfrac{\partial T}{\partial x} (0,t) = 0, \qquad  0<t<t_m.
\end{equation}
If heat transfer at the top surface is significant; i.e., there is any additional heat source at the top, Newton's law of cooling can be applied with a similar strategy described for the top surface.}

The initial temperature of the frozen region {\color{black} is assumed to be} spatially uniform at $T_0$,
\begin{equation} \label{eq:initial_temp}
    T(x,0) = T_0,  \qquad  0\leq x \leq L.
\end{equation}
{\color{black} The model assumes that sublimation does not begin until a sublimation temperature is reached at the top surface of the frozen material.} Therefore, we define the criterion for switching from the heating stage to the sublimation stage as
\begin{equation} \label{eq:switching_time}
   T(0,t_m) = T_m, 
\end{equation}
where $T_m$ is the sublimation temperature.

For the sublimation stage, the evolution of the sublimating interface (moving interface) is governed by the energy balance
\begin{equation} \label{eq:moving_interface}
    \dfrac{ds}{dt} = \dfrac{H_b(t)+H_{v2}L}{(\rho-\rho_d)\Delta H_{\textrm{sub}}}, \qquad t > t_m,
\end{equation}
where $s(t)$ is the interface position, $H_b(t)$ is the heat transfer from the bottom shelf, $H_{v2}$ is the microwave irradiation for the sublimation stage, $\rho_d$ is the density of the dried region, and $\Delta H_{\textrm{sub}}$ is the latent heat of sublimation. For the bottom shelf, 
\begin{equation}
    H_b(t) = h(T_b(t)-T). \label{eq:shelf_heat}
\end{equation}
During phase change, the temperature of the product is approximately uniform and constant at the sublimation point $T_m$ as the supplied heat is used only for sublimation \cite{Richard_2021_MainModel}, which is
\begin{equation}
    T = T_m, \qquad t > t_m. \label{eq:T_CFD}
\end{equation}
With the presence of microwave irradiation, experimental data in past publications show that the product temperature slightly increases during sublimation \cite{Gitter_2018_Experiment,Gitter_2019_Parameters}, which could be because some of the microwave irradiation, in addition to going into sublimation, can interact and heat the frozen material. Hence, the temperature of the frozen material is modeled by
\begin{equation}
    T(t) = T_m + \dfrac{H_{v3}}{\rho C_P}(t-t_m), \qquad t > t_m, \label{eq:T_MFD}
\end{equation}
where $H_{v3}$ is the microwave irradiation heating the product during sublimation. Without microwave irradiation (CFD), $H_{v3} = 0$, and thus \eqref{eq:T_MFD} reduces to \eqref{eq:T_CFD}. The initial interface position is
\begin{equation} \label{eq:initial_interface}
    s(t_m) = 0.
\end{equation}
To correlate the microwave irradiation with the actual power of the microwave, the power density $Q_v$ is defined as
\begin{equation} \label{eq:powerdensity}
    Q_v = \dfrac{Q}{V},
\end{equation}
where $Q$ is the output power of the microwave and $V$ is the volume of the product. As a result, the values of $H_{v1}, H_{v2}$, and $H_{v3}$ can be estimated by
\begin{gather} 
    H_{v1} = p_1Q_v, \\
    H_{v2} = p_2Q_v, \\
    H_{v3} = p_3Q_v,
\end{gather}
where $p_1$, $p_2$ and $p_3$ are the portions of the microwave power absorbed by the product. This set of parameters can be estimated from experimental data. This model is applicable to all modes of freeze drying. For CFD, $Q= 0$, and hence $H_{vi} = 0,$ for $i=1,2,3$. For MFD, $h = 0$, resulting in no heat conduction from the shelf.


\section{Modeling of Thermal Radiation} \label{sec:ThermalRadiation}
The technique for thermal radiation analysis used in this work is based on the diffuse gray surface model and radiation network approach, which is a well-known and reliable approach for modeling radiation exchange between multiple surfaces \cite{Mills_1995_HeatTransfer,Incropera_2007_HeatTransfer}. {\color{black} In this work, the technique will be implemented to the mechanistic model presented in Section \ref{sec:Model}; however, it is important to note that the technique itself can be similarly applied to other different mechanistic models for primary drying in the literature.}


\subsection{Derivation of the diffuse gray surface model} \label{sec:DiffuseGray}
The derivation of the diffuse gray surface model in this section is adapted from \cite{Mills_1995_HeatTransfer}. First consider the simplest case where a single vial is surrounded by the chamber wall without the presence of other vials (Fig.\ \ref{fig:SimpleGeometry}a); under this condition, the effect of thermal radiation is largest. The result from this derivation can be extended to analyze thermal radiation exchange between multiple surfaces in the later sections.

The vial surface and chamber wall in Fig.\ \ref{fig:SimpleGeometry}a are labeled as surface 1 and surface 2, respectively. At the surface of interest, all the radiation leaving that surface is defined as the radiosity, $J$ (W/m$^2$), whereas all the radiation arriving at that surface is defined as the irradiation, $G$ (W/m$^2$). Hence, the net radiative heat flux leaving surface 1 is  
\begin{equation} \label{eq:q1}
    q_1 = J_1 - G_1.
\end{equation}
All the radiation leaving surface 1 comprises the emitted and reflective components, which is given by
\begin{equation} \label{eq:J1}
    J_1 = \varepsilon_1\sigma T_1^4 + (1-\varepsilon_1)G_1,
\end{equation}
where $T_1$ is the temperature of surface 1, $\varepsilon_1$ is the emissivity of surface 1, and $\sigma$ is the Stefan-Boltzmann constant. 
Rearranging \eqref{eq:q1} and \eqref{eq:J1} gives
\begin{equation} \label{eq:q1J1}
    q_1 = \dfrac{\varepsilon_1}{1-\varepsilon_1}(\sigma T_1^4 - J_1).
\end{equation}
Hence, the net thermal radiation leaving surface 1 is 
\begin{equation} \label{eq:Q1}
    Q_1 = \dfrac{\varepsilon_1A_1}{1-\varepsilon_1}(\sigma T_1^4 - J_1),
\end{equation}
where $A_1$ is the area of surface 1. The vial can be assumed to be a cylinder, and so $A_1$ can be calculated from the vial diameter $d$ and product height $L$. The exact same analysis for surface 2, denoted by the subscript 2, results in
\begin{equation} \label{eq:q2J2_original}
    Q_2 = \dfrac{\varepsilon_2A_2}{1-\varepsilon_2}(\sigma T_2^4 - J_2).
\end{equation}
Next, consider the heat exchange between both surfaces. The radiant energy that leaves surface 1 and is intercepted by surface 2 is $J_1A_1F_{1-2}$, whereas the radiant energy that leaves surface 2 and is intercepted by surface 1 is $J_2A_2F_{21}$, where $F_{1-2}$ and $F_{2-1}$ are the view factors (aka shape factors). Therefore, the net radiative heat exchange from surface 1 to surface 2 is
\begin{equation} \label{eq:Q12_original}
    Q_{\textrm{rad}} = J_1A_1F_{1-2} - J_2A_2F_{2-1}.
\end{equation}
Application of the reciprocal rule gives that
\begin{equation} \label{eq:reciprocal}
    A_1F_{1-2} = A_2F_{2-1},
\end{equation}
which can be used to write \eqref{eq:Q12_original} as
\begin{equation} \label{eq:Q12}
    Q_{\textrm{rad}} = A_1F_{1-2}(J_1-J_2).
\end{equation}
Energy conservation requires that
\begin{equation} \label{eq:energy_balance}
    Q_1 = Q_{\textrm{rad}} = -Q_{2}.
\end{equation}
Combining \eqref{eq:q1J1}, \eqref{eq:q2J2_original}, \eqref{eq:Q12}, and \eqref{eq:energy_balance}, the final expression for $Q_{\textrm{rad}}$ is
\begin{equation} \label{eq:Q12_final}
    Q_{\textrm{rad}} = \dfrac{\sigma(T_1^4- T_2^4)}{\dfrac{1-\varepsilon_1}{\varepsilon_1A_1} + \dfrac{1}{A_1F_{1-2}} + \dfrac{1-\varepsilon_2}{\varepsilon_2A_2}}.
\end{equation}
In \eqref{eq:Q12_final}, the denominator can be viewed as a resistance to the thermal radiation. For convenience, define that resistance as 
\begin{equation} \label{eq:R_rad}
    R_{\text{rad}} =\dfrac{1-\varepsilon_1}{\varepsilon_1A_1} + \dfrac{1}{A_1F_{1-2}} + \dfrac{1-\varepsilon_2}{\varepsilon_2A_2}.
\end{equation}
For different geometry/configuration, the expression of $R_{\text{rad}}$ is varied. As suggested by \cite{Mills_1995_HeatTransfer,Incropera_2007_HeatTransfer}, it is useful to view this $R_\textrm{rad}$ for the single-vial case using an electrical network analogy as shown in Fig.\ \ref{fig:NetworkDefault}. The term $(1-\varepsilon)/\varepsilon A$ is usually defined as a surface resistance, whereas $1/A_1F_{1-2}$ is a space resistance. The surface resistance is dependent on the properties of that surface, and so this term is not influenced by other surfaces. On the other hand, the space resistance depends on the properties of the pair of surfaces, i.e., the view factor. Hence, when there are more than two surfaces, thermal radiation exchange between multiple surfaces is related to this space resistance. With this analogy, extension to radiation exchange between multiple surfaces can be done simply by adding more resistances into the network, which is shown in Section \ref{sec:MultipleSurfaces}.

For the single-vial case (two surfaces), \eqref{eq:Q12_final} can be simplified further. Originally, $T_1$ is the temperature of the vial surface. Nevertheless, as the model is simulated in one spatial direction, $T_1 $ becomes the temperature of the frozen material, which is $T(x,t)$. The wall temperature is assumed to be constant at $T_2$. Given the geometry shown in Fig.\ \ref{fig:SimpleGeometry}a, the radiant energy leaving surface 1 is all intercepted by surface 2; thus, the view factor $F_{1-2}$ is 1. As a result, \eqref{eq:Q12_final} becomes
\begin{equation} \label{eq:Q12_final_lyo}
    Q_{\text{rad}}(x,t) = \dfrac{\sigma ((T(x,t))^4-T_2^4)}{\dfrac{1-\varepsilon_1}{\varepsilon_1A_1} + \dfrac{1}{A_1} + \dfrac{1-\varepsilon_2}{\varepsilon_2A_2}},
\end{equation}
which has the resistance 
\begin{equation} \label{eq:R_rad_1vial}
    R_{\text{rad}} =\dfrac{1-\varepsilon_1}{\varepsilon_1A_1} + \dfrac{1}{A_1} + \dfrac{1-\varepsilon_2}{\varepsilon_2A_2}.
\end{equation}
With the net thermal radiation established, the energy conservation equation for the frozen region becomes
\begin{equation} \label{eq:governing_original2}
    \rho C_p\dfrac{\partial T}{\partial t} = k\dfrac{\partial^2 T}{\partial x^2} + H_v - \dfrac{Q_\textrm{rad}(x,t)}{V}, \qquad  0 < x < L, \qquad 0<t<t_m,
\end{equation}
where $V$ is the volume of the product. For the sublimation stage, the energy balance at the moving interface is
\begin{equation} \label{eq:moving_interface2}
    \dfrac{ds}{dt} = \dfrac{H_b(t)+H_vL - \frac{Q_\textrm{rad}(t)L}{V}}{(\rho-\rho_d)\Delta H_{\textrm{sub}} } , \qquad t > t_m.
\end{equation}
Other equations including the boundary and initial conditions remain the same. 

Although the above implementation is based on the mechanistic model presented in Section \ref{sec:Model}, the diffuse gray surface model can be similarly applied to any other models for freeze drying.


\subsection{View factor calculation} \label{sec:ViewFactor}
The view factor $F_{i-j}$, defined as the fraction of the radiant energy leaving surface $i$ that is received by surface $j$, is an important parameter that governs the significance of thermal radiation in the system. Hence, obtaining an accurate value of the view factor between surfaces is crucial. In the single-vial case (Section \ref{sec:DiffuseGray}), calculating the view factor is simple as there are only two surfaces, i.e., $F_{1-2} = 1$. For multiple vials, view factors can be calculated by using the analytical expressions \cite{Pikal_2016_Convection}, numerical integration \cite{Jiang_2020_NumerViewFactor}, the Monte Carlo method \cite{Mirhosseini_2011_MonteCarlo}, and 
estimation from experimental data \cite{Bano_2020_optimize}. 

In this work, we describe two techniques for determining view factors that can be implemented easily for the freeze-drying process. The first technique employs analytical expressions, which results in the exact value of the view factor. This method is straightforward but possible for simple systems where analytical expressions are available. The second approach relies on the Monte Carlo method, which is more complicated and computationally expensive than the analytical expression approach, but can be applied to any complicated geometry. 


\subsubsection{Analytical solutions}
Analytical solutions are available only for some simple geometries. View factors calculated from the analytical solutions are exact, and thus can be used to validate results obtained from more complex techniques such as the Monte Carlo method. Here we consider two simple cases where the analytical solutions are obtained from \cite{Mills_1995_HeatTransfer}. 

The first case consists of two vials in the chamber as shown in (Fig.\ \ref{fig:SimpleGeometry}b), where $c$ is the distance between vials, the left vial is denoted as $1l$, and the right vial is denoted as $1r$. Due to symmetry, thermal radiation from the chamber wall should affect both vials equally, and hence both vials have the same temperature at all times. As a result, both vials are identical, and the view factor can be calculated analytically by
    \begin{equation} \label{eq:F_1lr2}
        F_{1l-2} = F_{1r-2} = 1 - \dfrac{1}{\pi}\!\left(\sqrt{Y^2-1} + \sin^{-1}\!\left(\dfrac{1}{Y}\right) - Y \right)\!,
    \end{equation}

where $Y = 1 + c/d$. 

Another case follows the geometry in Fig.\ \ref{fig:SimpleGeometry}c, where there are three vials: the left vial is denoted as $1l$, the right vial is denoted as $1r$, and the middle vial is labeled as $1m$. In this case, $F_{1l-2}$ and $F_{1r-2}$ can also be calculated using \eqref{eq:F_1lr2}. For $1m$, since it is surrounded by two identical vials,
    \begin{equation} \label{eq:F_1m2}
        F_{1m-2} = 1 - \dfrac{2}{\pi}\!\left(\sqrt{Y^2-1} + \sin^{-1}\!\left(\dfrac{1}{Y}\right) - Y \right)\!.
    \end{equation}
{\color{black} The view factors obtained from the analytical solutions for the two-vial problem (three surfaces) and three-vial problem (four surfaces) presented here are only used for validating the Monte Carlo simulation that are presented in the next section. In other cases, the Monte Carlo method is used.}


\subsubsection{Monte Carlo simulation} \label{sec:MonteCarlo}
The main advantage of the Monte Carlo method is that it provides a systematic framework for calculating the view factor for any complex geometry. The drawback is that, when the number of surfaces is significantly high, the Monte Carlo method can be computationally intensive. Nevertheless, the view factor is not required to be calculated online or in real-time when the freeze dryer is being operated; the view factor can be computed when the vial disposition is known during the design stage. 

In this work, the Monte Carlo simulation is implemented in MATLAB, which can be summarized in three steps. First, the geometry and vial disposition in the chamber are defined, which is done by using a set of coordinates in a 2D plane. Second, a number of rays are shot randomly from the surface of interest to represent thermal radiation using \texttt{rand}. Alternatively, these rays can be uniformly placed, but random shooting usually gives a better convergence given the same number of iterations. Finally, the view factor $F_{1-2}$ can be obtained by calculating the ratio of the number of rays intercepted by surface 2 to the number of rays shot from surface 1. Interception is indicated by intersection between curves, which is computed using the algorithm in \cite{Schwarz_2023_intersect}. This algorithm is based on a 2D plane, meaning that the cylinder (vial) is assumed to have infinite length. The error of this approximation on the view factor is tiny provided that (1) the distance between vials ($c$) is small and {\color{black} (2) the vial height} is large compared to the radius of the vial \cite{Juul_1982_FiniteCylinder}. The former is generally true for freeze drying of multiple vials, while the latter is common for typical vials.


\subsection{Thermal radiation exchange between multiple surfaces} \label{sec:MultipleSurfaces}
In practice, there are many vials in a freeze dryer, especially in pharmaceutical and biopharmaceutical manufacturing \cite{Veraldi_2008_Parameters,Bano_2020_optimize,Bhambhani_2021_MVD}, and so the framework introduced in Section \ref{sec:DiffuseGray} needs to be modified. Previous studies modeled thermal radiation in freeze drying by assuming that radiation exchange exists between each vial and the chamber wall independently (two surfaces) \cite{Gan_2005_WallTemp,Veraldi_2008_Parameters,Pikal_2016_Convection,Bano_2020_optimize}. In reality, thermal radiation exchange exists not only between the vial and chamber wall but also between multiple vials, where all of these processes occur simultaneously. To our knowledge, there is no literature discussing a systematic way of modeling this complicated behavior in the context of freeze drying. In this article, we adapt the radiation network approach discussed in \cite{Mills_1995_HeatTransfer,Incropera_2007_HeatTransfer} to describe radiation exchange between multiple surfaces. The section firstly discusses a complete network representation technique and then proposes some simplified techniques specifically for freeze drying.


\subsubsection{Radiation network approach} \label{sec:RadiationNetwork}
The radiation network is briefly introduced in Fig.\ \ref{fig:NetworkDefault} for the two-surface case. Before discussing the technique for multiple surfaces, it is important to note two critical relations for radiation exchange between $k$ surfaces:
\begin{gather}
    \sum_{j=1}^k F_{i-j} = 1, \label{eq:summation} \\ 
    A_iF_{i-j} = A_jF_{j-i}. \label{eq:recip}
\end{gather}
These two equations describe the relations between view factors and surface area, which are valid for all pairs of surfaces in the network.

To demonstrate the radiation network technique, consider the three-vial case (Fig.\ \ref{fig:SimpleGeometry}c). In this case, there are four surfaces, where radiation exchange exists between (1) surfaces $1l$ and 2, (2) surfaces $1m$ and 2, (3) surfaces $1r$ and 2, (4) surfaces $1l$ and $1m$, and (5) surfaces $1r$ and $1m$. The network representation of the four surfaces is illustrated in Fig.\ \ref{fig:Network4Surfaces}.

Each pair of two surfaces is connected via the space resistance. The key idea of the network representation technique is to ensure the radiant energy balance holds for all surfaces in the system. For example, by considering node $J_2$, the radiant energy balance can be described by
\begin{equation}
    \dfrac{\sigma T_2^4-J_2}{(1-\varepsilon_2)/\varepsilon_2A_2} = \dfrac{J_2-J_{1l}}{1/A_{1l}F_{1l-2}} + \dfrac{J_2-J_{1m}}{1/A_{1m}F_{1m-2}} + \dfrac{J_2-J_{1r}}{1/A_{1r}F_{1r-2}},
\end{equation}
which uses the convention that heat transfers from surface 2 to the other surfaces. 
The same analysis can be performed for the other three nodes, namely $J_{1l}$, $J_{1m}$, and $J_{1r}$. In primary drying, the properties and temperature of each surface are known (e.g., from the initial conditions), so the only unknowns are $J_{1l}$, $J_{1m}$, $J_{1r}$, and $J_{2}$. Here we have a linear system of four equations resulting from the energy balance equations and four unknowns, and hence the system can be solved efficiently and straightforwardly, e.g., using \texttt{mldivide} (backslash) in MATLAB. For $k$ surfaces, the energy balance equation can be written as
\begin{equation}
    J_i = \varepsilon_i \sigma T_i^4 + (1-\varepsilon_i)\sum_{j=1}^{k} J_i F_{i-j}, \qquad \textrm{for} \ i = 1,2,...,k,
\end{equation}
which results in a linear system of $k$ equations and $k$ unknowns. When $J_i$ is calculated, the net radiant energy from each surface $i$ can be obtained by
\begin{equation}
    Q_{\textrm{rad},i} = \dfrac{\varepsilon_iA_i}{1-\varepsilon_i}(\sigma T_i^4 - J_i),
\end{equation}
where $Q_{\textrm{rad},i}$ replaces $Q_{\textrm{rad}}$ in the freeze-drying model given by \eqref{eq:governing_original2} and \eqref{eq:moving_interface2}.

To incorporate this framework into the dynamic modeling of primary drying, the first step is to initialize the mechanistic model \eqref{eq:governing_original2} and \eqref{eq:moving_interface2} for all vials, i.e., one model for one vial. Then, solve the radiation network to calculate $Q_{\textrm{rad},i}$ for every vial simultaneously. Finally, the equation is numerically integrated to the next time step. The calculation procedure is summarized in Fig.\ \ref{fig:Flowchart_combined}a.

This radiation network approach has several benefits. First, radiation exchanges between every surface are captured accurately. Secondly, the approach ensures the conservation of radiant energy in the system. Lastly, the approach can be systematically applied to model any complicated freeze-dryer design and vial disposition regardless of the number of surfaces. The only drawback is that the network representation could be highly complex when the number of vials is high, which is quite common in industrial freeze dryers. The number of partial differential equations (PDEs) and complexity of the linear system of equations are dependent on the number of vials, thereby intensive computation. Although that does not prohibit the use of this radiation network approach, some approximation/simplification, which results in much faster computation, is discussed in the next section. 


\subsubsection{Simplified approach} \label{sec:SimNetwork}
There are different ways to simplify the radiation network approach for primary drying in freeze drying. The simplest technique for freeze drying assumes that the interaction between vials is negligible and all vials are independent, and so thermal radiation exists only between each vial and the chamber wall, i.e., two surfaces at a time. This technique is common in literature due to its simplicity. In such cases, the net radiant energy exchange between each vial, denoted as $i$, and the chamber wall, denoted as 2, is approximated by
\begin{equation} \label{eq:Qrad_sim}
    Q_{\textrm{rad},i} = \dfrac{\sigma(T_{i}^4- T_2^4)}{\dfrac{1-\varepsilon_{i}}{\varepsilon_{i}A_{i}} + \dfrac{1}{A_{i}F_{i-2}} + \dfrac{1-\varepsilon_2}{\varepsilon_2A_2}},
\end{equation}
where $Q_{\textrm{rad},i}$ replaces $Q_{\textrm{rad}}$ in the freeze-drying model given by \eqref{eq:governing_original2} and \eqref{eq:moving_interface2}. 

The calculation procedure for the simplified technique is given in Fig.\ \ref{fig:Flowchart_combined}b. The simplification decouples the radiation network; i.e., all vials are independent, and so there is no linear system of equations to be solved. Also, the simplified approach can be selectively applied to the vial of interest, whereas the radiation network approach requires simultaneous modeling of all vials as they are coupled in the network. This simplification can significantly lower the computational cost.

The simplified approach is a good approximation for primary drying in freeze drying for two reasons. First, the temperature of the vial is nearly constant at the sublimation temperature most of the times. Second, the material and size of all vials are the same. Therefore, these vials are approximately identical, which justifies that radiation exchange between the vials should be relatively small compared to that between the vials and chamber wall. Although this technique results in a much simpler calculation and faster simulation, significant error could occur in some cases, which is discussed further in Section \ref{sec:ModelCompare}. 


\subsubsection{Hybrid approach and parameter estimation} \label{sec:ParaEst}
The radiation network and simplified approaches described above do not require any parameter estimation or fitting for the radiation component. Instead, all relevant parameters can be calculated analytically or numerically. Nevertheless, our framework also provides flexibility for parameter estimation from data. In such cases, we rely on the formulation of the simplified approach by assuming that thermal radiation exchange exists independently between each vial and the chamber wall. However, instead of using \eqref{eq:Qrad_sim}, the radiative heat flux is expressed by
\begin{equation} \label{eq:Q12_exp}
     Q_{\textrm{rad},i} = \dfrac{\sigma (T_i^4-T_2^4)}{R_\textrm{rad}},
\end{equation}
where $R_\textrm{rad}$ is the resistance to thermal radiation estimated from data, e.g., drying time. Also, in the case of unknown wall temperature, its value $T_2$ can be estimated from data. When $R_\textrm{rad}$ is obtained, the modeling procedure is identical to that of the simplified approach. Data for parameter estimation could be experimental data or data obtained from the radiation network approach.

This hybrid approach relies on a combination of data and first-principles modeling. It can provide a highly accurate result when calibrated by data, but is specific to the system that the data have been collected from. By having its computational cost the same as the simplified approach, the hybrid model can be computed much faster than for the radiation network approach.  Another advantage is that the hybrid model does not require the calculation of view factors because those parameters are included in $R_\textrm{rad}$, which is estimated from data.



\subsection{Summary of model implementation and limitation}
The original model presented in Section \ref{sec:Model} can be solved analytically or numerically; we refer to the detailed procedure in \cite{Srisuma_2023_AnalyticalLyo}. For thermal radiation analysis, the view factor should be calculated first using the Monte Carlo method described in Section \ref{sec:MonteCarlo}. For the radiation network approach, the model integrated with the radiation network should be solved numerically, which follows the calculation procedure shown in Fig.\ \ref{fig:Flowchart_combined}a. For the simplified approach, the calculation procedure follows Fig.\ \ref{fig:Flowchart_combined}b. In general, we recommend parameter estimation for the heat transfer coefficient ($h$), microwave power distribution ($p_1, p_2, p_3$) for MFD/HFD, and sublimation temperature ($T_m$) as these parameters can vary greatly among systems. Parameter estimation for the radiation part is not necessary but possible as with the hybrid model explained in Section \ref{sec:ParaEst}.

The main advantage of our model is that it is not limited to the number of surfaces or geometry. Therefore, this modeling strategy can be systematically applied to analyze complicated freeze-dryer design regardless of the number of vials or geometry. The radiation network approach relies on three assumptions \cite{Mills_1995_HeatTransfer}: (1) all surfaces are opaque and gray, (2) the emission and reflection from all surfaces are diffuse, and (3) the radiosity of each surface is uniform. The first two assumptions are valid in most engineering applications. The third assumption is not exactly true for some systems; for example, the radiosity near the vertex of the chamber wall does not have to be equal to the radiosity at the center due to asymmetry. Nevertheless, this error can be reduced by dividing a surface into smaller surfaces of acceptably uniform radiosity, where the number of smaller surfaces depends on the level of accuracy needed \cite{Mills_1995_HeatTransfer}. Since the general framework of the radiation network is not limited by the number of surfaces, adding more surfaces to the system is not an issue.

{\color{black} Since the mechanistic model in Section \ref{sec:Model} considers sublimation as a heat transfer-controlled process, the product temperature is approximately constant during the sublimation stage for CFD. Hence, the computed effect of thermal radiation on the product temperatures is negligible for the sublimation stage. The proposed framework for modeling thermal radiation is not limited to any specific mechanistic model, however, and can be applied to situations where mass transfer is important.}

Along with this article, we provide the MATLAB implementation of our model (see Section \ref{sec:Code}), which includes the original mechanistic model, the radiation network model for thermal radiation analysis, and the Monte Carlo simulation for view factor calculation. Users can freely set the inputs to simulate their systems of interest. Examples of model implementation are shown in Section \ref{sec:Analysis}.


\section{Simulation and Model Validation} \label{sec:ModelValiation}
The mechanistic model and several modeling strategies discussed in Sections \ref{sec:Model} and \ref{sec:ThermalRadiation} are validated using simulation studies and experimental data from the literature. The default parameters are listed in Table \ref{Tab:Parameters}; parameter values different from those reported in the table are stated explicitly in that specific section.


\subsection{Validation of the Monte Carlo method}
In this work, the Monte Carlo method is employed for view factor calculation as it can flexibly handle different geometry and vial disposition. Before applying the Monte Carlo method, however, it is important to ensure that the algorithm provides an accurate prediction of the view factor. This section compares view factors calculated from the Monte Carlo simulation with the analytical solutions for some simple vial layouts, namely, for two and three vials (Figs.\ \ref{fig:SimpleGeometry}bc) with the default parameters in Table \ref{Tab:Parameters}.

Table \ref{Tab:MonteCarlo} shows that the Monte Carlo method can provide highly accurate prediction of the view factors for the layouts considered in  Figs.\ \ref{fig:SimpleGeometry}bc. The error of calculation is on the order of $0.1\%$, which is practically negligible. As such, this method is promising for use in the calculation of view factors in our freeze-drying system.


\subsection{Comparison of modeling strategies} \label{sec:ModelCompare}
Different techniques for modeling radiation exchange between multiple surfaces are discussed in Section \ref{sec:MultipleSurfaces}. Here we compare the radiation network approach with the simplified approach to justify its approximation. The former considers radiation exchange between all surfaces simultaneously and accurately, while the latter assumes that thermal radiation exists only between each vial and the chamber wall independently. 

Consider a rectangular array of 10$\times$10 vials with the default parameters in Table \ref{Tab:Parameters}. Using the radiation network approach, the drying times of all vials can be obtained as shown in Fig.\ \ref{fig:Vial_10by10_Comparison}a. Due to the locations, the four corner vials have the highest view factor as they are exposed to the chamber wall more than any other vials. The corner vials dry the fastest, at about 9.6 hours, as the influence from thermal radiation is largest. The second group of vials that is dried is the edge vials, with drying time of about 11.5 hours. The effect of thermal radiation is largest for the outermost vials and becomes significantly weaker for inner vials. The error of using the simplified method is about 5\%--7\% for the outermost vials (Fig.\ \ref{fig:Vial_10by10_Comparison}b). {\color{black} This error is equivalent to about 0.7 hours of drying time, indicating that the radiation exchange between vials can be significant.}

The radiation network approach captures all radiation exchanges between multiple vials and the chamber wall, and thus provides physically reasonable and accurate results. Nevertheless, the radiation network approach is much more computationally expensive than the simplified approach, in particular when the number of vials is high. For 100 vials, the simulation time of the radiation network approach is on the order of several minutes, whereas for the simplified approach is on the order of seconds. As such, the simplified approach can be useful in applications where speed is crucial, e.g., real-time/online simulation, but the error of simplification should be quantified properly on a case-by-case basis, with respect to the radiation network approach. In Section \ref{sec:ML}, we demonstrate the use of the hybrid approach, which combines the simplified technique and the radiation network representation for fast simulation with high accuracy.


\subsection{Model validation with simulation and experimental data}
The mechanistic model is validated using various simulation and experimental data, ranging from cases for internal vials (i.e., no radiation) to edge and corner vials with thermal radiation. The parameters specific to each case study are given in Table \ref{Tab:ParametersSpecific}, with other parameters following the default values in Table \ref{Tab:Parameters}.


\subsubsection{Conventional freeze drying} \label{sec:CFD}

For CFD, simulation results are compared with two sets of experimental data from the literature, in which the temperature was reported for inner vials surrounded by outer vials, so that the effects of thermal radiation are minimal.\footnote{The outer vials act as a thermal radiation shield.} 

The first data set, denoted as Case 1, is obtained from \cite{Gitter_2018_Experiment}. Our model is able to simulate the product temperature at the bottom and predict the drying time accurately compared with the experimental data (Fig.\ \ref{fig:Case1}). The temperature is nearly constant during sublimation, indicating a heat transfer-controlled process. The only difference is at the transition between the heating and sublimation stages, where the experimental data show a smoother transition. 

Another set of data, denoted as Case 2, is obtained from \cite{Veraldi_2008_Parameters}, where the interface position profiles are available. In this case, two experiments were conducted at different shelf temperatures: 258.15 and 268.15 K. 

Our model is able to predict the drying time, product temperature, and interface position in both cases reasonably well (Figs.\ \ref{fig:Case2_1} and \ref{fig:Case2_2}). The only significant difference is that the measured bottom temperatures slightly increase over time, which implies some contribution of mass transfer. However, ignoring mass transfer does not significantly impact the ability of our model to identify the end point of primary drying. At the end of primary drying, the measured temperature appears to increase more abruptly, indicating the start of secondary drying.


\subsubsection{Microwave-assisted freeze drying} \label{sec:MFD}

Experimental data for MFD are relatively limited. Here we use two sets of data from \cite{Gitter_2018_Experiment} and \cite{Gitter_2019_Parameters}, denoted as Cases 3 and 4, respectively. The temperature was reported for inner vials surrounded by shielding vials, so that the effects of thermal radiation are minimal. 

The model can be used to estimate the evolution of the product temperature in MFD accurately (Fig.\ \ref{fig:Case3}). Correspondingly, the drying time can be predicted. An important observation is that the microwave power distribution, i.e., $p_1,p_2,p_3$, varies significantly among different experiments. This is because microwave absorption is influenced by several factors in a complicated manner, e.g., solute concentration and microwave frequency \cite{Bhambhani_2021_MVD}, and so these microwave-related parameters are usually estimated from data.\cite{Richard_2021_MainModel}. 

For Cases 1 to 4, we demonstrate that our model can be used to simulate the product temperature and interface position during primary drying for both CFD and MFD (HFD is just a combination of both) without thermal radiation, i.e., for inner vials. Model validation in the next section focuses on edge/corner vials, where thermal radiation is significant.


\subsubsection{Effects of thermal radiation} \label{sec:ThermalRadValidation}

{\color{black} Model validation in Sections \ref{sec:CFD} and \ref{sec:MFD} considers data for inner vials, where the effect of thermal radiation is negligible, to validate our base model in Section 2. This section focuses on the validation of the thermal radiation model (Section \ref{sec:ThermalRadiation}).}

The first set of data for thermal radiation, denoted Case 5, is obtained from the simulation results of \cite{Veraldi_2008_Parameters}, where CFD was considered. The wall temperature $T_2$ and view factor $F_{1-2}$ were not reported. Thus, we estimate $T_2$ from data and calculate $F_{1-2}$ using the Monte Carlo approach. In \cite{Veraldi_2008_Parameters}, radiation exchange was calculated between the vial and chamber wall only, so we use the simplified approach here for a fair comparison.

From Table \ref{Tab:Result_Case4}, {\color{black} the predicted drying time is lower than the reference data by about 0.3 hours (4\%) for our default wall temperature of 293.15 K}. This error can be reduced by estimating the wall temperature from data. The estimated wall temperature of 288.80 K lies in a typical range found in the literature \cite{Gan_2005_WallTemp,Pikal_2016_Convection,Bano_2020_optimize}. An important observation here is that wall temperature plays an important role in thermal radiation. Hence, this parameter should be accurately measured or estimated from data rather than relying on some default value from the literature. This wall temperature effect is discussed in more detail in Section \ref{sec:WallTemp}.

Another set of data is from the experiments of \cite{Gan_2005_WallTemp}, where CFD was considered. In this experiment, most thermophysical properties were not reported, so the default values in Table \ref{Tab:Parameters} are used. The heat transfer coefficient is estimated using the data of the center vial (no radiation). The radiation network technique is applied without any parameter estimation for the radiation component to demonstrate the robustness of thermal radiation modeling.

The predicted drying times agree quite well with the experimental data, with the maximum error of about 3.5\% (Table \ref{Tab:Result_Case5}) for the corner vial. This error is relatively small given that there is no parameter estimation for the thermal radiation part of the model, indicating the accuracy of the radiation network approach.\footnote{The error would be larger if the simplified approach was used because the simplified approach always underestimates the drying time as discussed in Section \ref{sec:ModelCompare}.}


\section{Analysis and Parametric Studies} \label{sec:Analysis}
In Section \ref{sec:ModelValiation}, our mechanistic model is able to accurately capture the dynamic changes in temperature and interface position during primary drying. Furthermore, the implementation of a radiation network in our model effectively addresses the influence of thermal radiation on the drying time, ensuring a proper consideration of this important factor. Here we showcase the applications of our model via a comprehensive analysis and parametric study of thermal radiation in primary drying. Results presented in this section are obtained from the radiation network approach, i.e., no simplification or approximation.


\subsection{Different modes of freeze drying} \label{sec:ThreeModes}
The influence of thermal radiation on different freeze-drying modes varies, with previous literature primarily concentrating on CFD. To gain insights into this phenomenon, we apply our mechanistic model incorporating the radiation network to predict the drying times of an array of 10$\times$10 vials for CFD, MFD, and HFD, with the default parameters in Table \ref{Tab:Parameters}.

From Fig.\ \ref{fig:ThreeModes}, CFD has the largest variation in drying times, ranging between 9.5 hours to 17.2 hours, while the variation is smallest in HFD, which ranges from 2.5 hours to 3.2 hours. This result is understandable because microwave irradiation plays a role in enhancing the uniformity of heat transfer in the system, which is a benefit of using MFD and HFD beyond drying time reduction \cite{Abdelraheem_2022_Uniformity}. In all cases, the inner vials are slightly influenced by thermal radiation, agreeing with experimental observations in the literature \cite{Fissore_2018_Review,Veraldi_2008_Parameters}.

{\color{black} In Table \ref{Tab:Heat_rad}, the corner vials receive a largest amount of radiative energy, whereas the value is much smaller for the center vials. In CFD, the total radiative energy is larger than those in MFD and HFD because the total drying time of CFD is much longer, and so there is more time for thermal radiation to occur. In MFD and HFD, the process is completed much faster due to the contribution of microwave irradiation, and so the contribution of thermal radiation is smaller. This analysis agrees with the physical interpretation of heat transfer and is consistent with the drying time results presented earlier.}


\subsection{Wall temperature} \label{sec:WallTemp}
In the literature, the wall temperature is usually higher than that of the vial, and thus the radiant energy from the wall accelerates the drying process \cite{Gan_2005_WallTemp,Ehlers_2021_WallTemp}. In practice, the wall temperature varies greatly among different systems because it depends on various factors, e.g., freezing steps, drying protocol, freeze-dryer design, and external environment, and thus the wall temperature needs to be measured or controlled \cite{Gan_2005_WallTemp,Pikal_2016_Convection,Bano_2020_optimize,Ehlers_2021_WallTemp}. Some past studies found that the wall temperature was relatively constant during the drying step \cite{Veraldi_2008_Parameters,Bano_2020_optimize}. 

Wall temperature plays an important role in thermal radiation as shown in Section \ref{sec:ThermalRadValidation}, and hence its effects are investigated here. {\color{black} The wall temperature is assumed to be constant throughout the drying process to simplify our analysis, although the radiation network approach is not restricted to this assumption.} The literature has reported several wall temperature values; the maximum value to our knowledge is 293.15 K reported by \cite{Gan_2005_WallTemp}, which is the default value in Table \ref{Tab:Parameters}. The wall temperature, $T_2$, is varied from the sublimation temperature of 256.15 K up to 293.15 K while keeping other parameters as in Table \ref{Tab:Parameters}. HFD is considered.

An increase in the wall temperature leads to a notable reduction in the time required for drying (Fig.\ \ref{fig:WallTemp_HFD}). If the wall temperature is equal to the sublimation temperature (the first data point on the left), the drying time is not significantly affected because the wall and vial temperatures are similar. At the highest wall temperature of 293.15 K, the drying time is decreased by about 0.61 hours for the corner vials and 0.43 hours for the edge vials, which corresponds to about 19\% and 14\% reduction, respectively. Therefore, it is crucial to carefully monitor and control the wall temperature to accurately assess the effects of thermal radiation.

\subsection{Vial disposition}
The number of vials and its deposition vary greatly among different freeze-drying systems, which directly influences the impact of thermal radiation. The radiation network framework proposed in this work enables the analysis of any complicated vial disposition and freeze-dryer design. For the analysis in this section, all parameters are based on the default values in Table \ref{Tab:Parameters} unless otherwise specified. HFD is considered.

Firstly, the six different vial layouts are investigated (Table \ref{Tab:Dispo_layout}). The drying time is longer as the number of vials increases, which is understandable because additional vials act as a radiation shield reducing the view factor between each vial and the chamber wall. The drying time exhibits minimal variation when the number of vials reaches some certain thresholds. This phenomenon arises from the fact that, once the number of vials surpasses a certain point, the addition of more vials has a negligible effect on the view factor of the existing vial. This insight is useful as it limits the number of vials to be modeled at this threshold instead of modeling every vial added to the system, reducing computational time.

Another important aspect of vial disposition is the distance/gap between vial, $c$. Here we compare the case where $c = 0.5$ cm (the default value as in Table \ref{Tab:Parameters}) with the case where there is no gap between vials, i.e., $c=0$ cm. The result shows that reducing the gap between vials prolongs the drying (Fig.\ \ref{fig:Dispo_distance}). When the vial gap is smaller, the vials are packed more closely. This results in a reduction in the view factor between the vials and chamber wall, and so reduces the effect of thermal radiation. Given that the drying time for the non-radiation case is 3.17 hours, thermal radiation has a slight heating effect on the inner vials when $c=5$ cm (Fig.\ \ref{fig:Dispo_distance}a) and has no effect on the inner vials for $c=0$ cm (Fig.\ \ref{fig:Dispo_distance}b). {\color{black} This analysis suggests that empty vials could be used as a radiation shield, particularly if the vials are packed closely, agreeing with the literature \citep{Hottot_2005_HTC,Veraldi_2008_Parameters,Gitter_2018_Experiment,Gitter_2019_Parameters}.}  In such cases, a more complicated technique might be needed. If properly shielded, all the inner vials can be modeled without considering thermal radiation. 

Lastly, our model can be applied to analyze different array structures. In freeze drying, two common arrangements for vial disposition are rectangular and hexagonal arrays \cite{Gan_2005_WallTemp,Hottot_2005_HTC,Veraldi_2008_Parameters,Pikal_2016_Convection}. By using 10$\times$10 vials with $c = 0$ cm in CFD as an example, Fig.\ \ref{fig:Dispo_CP} shows that the hexagonal array has a larger variation in the drying times compared to the rectangular array, which is understandable considering a more symmetric structure of the rectangular array. The hexagonal array, however, is the most optimal way for space management; i.e., given the same amount of space, more vials can be added to the hexagonal array as it represents the closest packed structure.

{\color{black} Our analysis focuses on thermal radiation from the chamber walls, which contributes to the majority of radiative heat transfer in the system due to its large surface area. Another source of thermal radiation can be the heating shelf, which affects all vials on the shelf irrespective of their locations. This effect is generally incorporated in \eqref{eq:boundary_bottom}, but can also be handled by the radiation network by considering the heating shelf as another object.}


\subsection{Training the hybrid model} \label{sec:ML}
As explained in Section \ref{sec:ModelCompare}, the simplified approach underestimates the drying time as it does not capture the radiation exchange between vials. However, the simplified technique has an advantage over the radiation network approach for applications requiring fast computation or real-time simulation. Here we demonstrate the application of the hybrid approach introduced in Section \ref{sec:ParaEst} to train the simplified model with the radiation network model. The resulting hybrid model combines the benefits of the radiation network and simplified approaches, leading to a fast and accurate simulation. In this case, we use the radiation network simulation as training data. The exact same procedure can be used for cases where experimental data is used for training.

Hybrid model training involves solving the optimization 
\begin{align}\label{eq:Training}
&\min_{R_\textrm{rad}} \ (t_\textrm{data} - t_\textrm{hybrid})^2 \\
&\textrm{s.t.} \nonumber \\
    &\textrm{Equations \eqref{eq:governing_original2}, \eqref{eq:moving_interface2}, \eqref{eq:Q12_exp}} \nonumber,
\end{align} 
where $t_\textrm{data}$ is the ground truth and $t_\textrm{hybrid}$ is the drying time predicted by the hybrid approach. The optimization \eqref{eq:Training} is solved independently for each vial of interest, which could be only some vials or every vial in the chamber. 

In this case, consider a rectangular array of 10$\times$10 vials in CFD, with the default parameters in Table \ref{Tab:Parameters}. The drying times predicted by the radiation network approach at $T_2 = 293.15$ K are used as the only training data, which corresponds to Fig.\ \ref{fig:Vial_10by10_Comparison}a. The hybrid model is trained for all 100 vials, resulting in the values of $R_\textrm{rad}$ shown in Fig.\ \ref{fig:HybridModel}a. The obtained values are logical; $R_\textrm{rad}$ is low for corner and edge vials, implying that these vials are significantly affected by thermal radiation, whereas $R_\textrm{rad}$ is high for inner vials as the effect of thermal radiation is insignificant there.

After $R_\textrm{rad}$ is obtained, the hybrid approach is tested with four test data, which are the drying times predicted by the radiation network approach at different wall temperatures as illustrated in Fig.\ \ref{fig:HybridModel}b. {\color{black} The error is less 0.01 hours in all cases, which is not practically different.} The hybrid approach is able to accurately predict the drying times for other wall temperatures given only one training set at 293.15 K, indicating the robustness of our model and training procedure. Logically, the accuracy of the hybrid approach can be improved with more training data.

{\color{black}
\subsection{Other applications} \label{sec:other}
Many applications of the radiation network approach to freeze drying are demonstrated in this article. This section briefly discusses other freeze-dryer designs that could be analyzed using the radiation network approach. 

As discussed before, the key benefit of this approach is that it can be systematically applied to model complicated freeze-dryer designs, irrespective of the number of objects and surfaces. For example, in batch freeze drying, it is common that vials are loaded using a stainless-steel tray or frame. The radiation network could be applied to analyze the effect of thermal radiation associated with this additional component. This stainless-steel frame becomes an additional object in the radiation network that partially shields the vials from the chamber walls. The view factors can be recalculated and so the same strategy described in this work can be employed. For example, a simulation for the HFD case in Section \ref{sec:ThreeModes} with the tray of height of 4.2 cm (same as the sample height) gives that the drying times of the corner and edge vials increase from 2.56 to 2.80 hr and from 2.74 to 2.92 hr, respectively. The presence of this additional object slightly reduces the effect of thermal radiation on the vials.   

Other possible scenarios that are not covered in this work but can be analyzed via our proposed framework include analysis of radiation shields, asymmetric chamber walls, and continuous freeze dryers \cite{Capozzi_2019_conlyo,Pisano_2019_conlyo}. In any case, the procedure is still the same: (1) define the geometry and surface, (2) calculate the view factors, and (3) apply the radiation network approach.
}


\section{Conclusion}
A new mechanistic model is proposed for primary drying in conventional (CFD), microwave-assisted (MFD), and hybrid (HFD) freeze drying. The model incorporates the diffuse gray surface model with a radiation network that accurately accounts for thermal radiation exchange between the vials and chamber wall. This mechanistic approach is the first that offers a systematic framework for simulating thermal radiation between multiple surfaces in any complex freeze-dryer designs. A simplification technique is also introduced, which trades off model accuracy with significantly faster computation. The hybrid approach allows for the combination of first-principles modeling and data to increase accuracy.

Our framework is extensively validated with analytical solutions, past simulation studies, and experimental data from the literature. The proposed model is able to accurately simulate the evolution of temperature and interface position during the primary drying phase. {\color{black} The model can effectively predict and assess the impacts of thermal radiation in freeze drying across different situations.} The key strength of our model lies in its comprehensive consideration of thermal radiation exchange between the vials and chamber wall. Consequently, the model enables accurate prediction of drying times for all vials, including corner vials, edge vials, and inner vials. We demonstrate  applications of the model for analyzing the impact of various parameters on thermal radiation in freeze drying, including the mode of operation (CFD, MFD, and HFD), wall temperature, and vial disposition. Our framework and analysis can be used to facilitate the design and optimization of new and existing freeze dryers.

\section*{Data Availability}  \label{sec:Code}
The MATLAB implementation of our mechanistic model, the radiation network, and Monte Carlo method for view factor calculation is available at \url{https://github.com/PrakitrSrisuma/Lyo-Radiation-Modeling}.

\section*{Acknowledgements} 
This research was supported by the U.S. Food and Drug Administration under the FDA BAA-22-00123 program, Award Number 75F40122C00200.

\clearpage

\begin{figure}[ht!]
    \centering
    \includegraphics[scale=0.5]{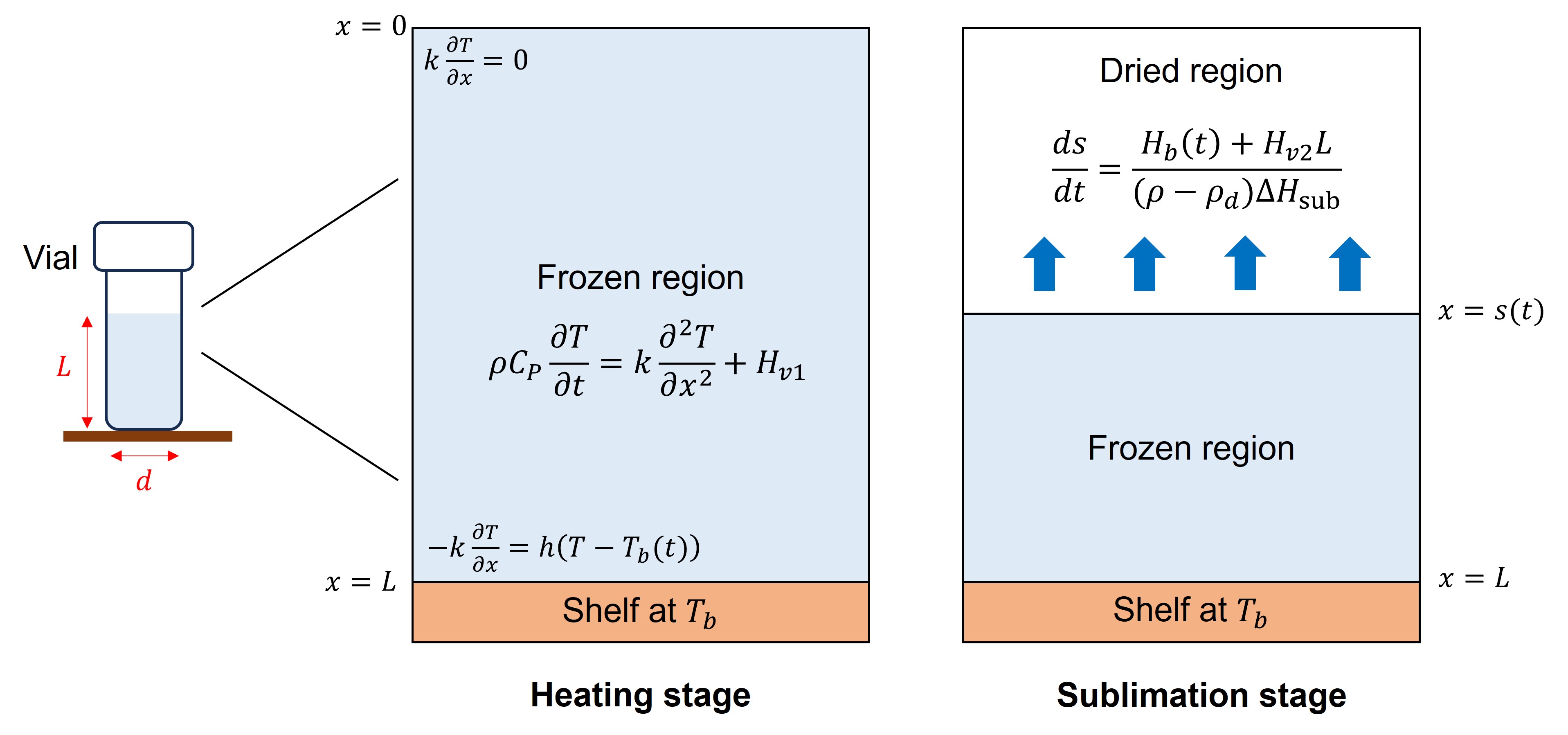}    
    \caption{Schematic diagram of the freeze-drying process. The figure is adapted from \cite{Srisuma_2023_AnalyticalLyo}.}
    \label{fig:Model_Schematic}
\end{figure}

\begin{figure}[ht!]
    \centering
\includegraphics[scale=0.5]{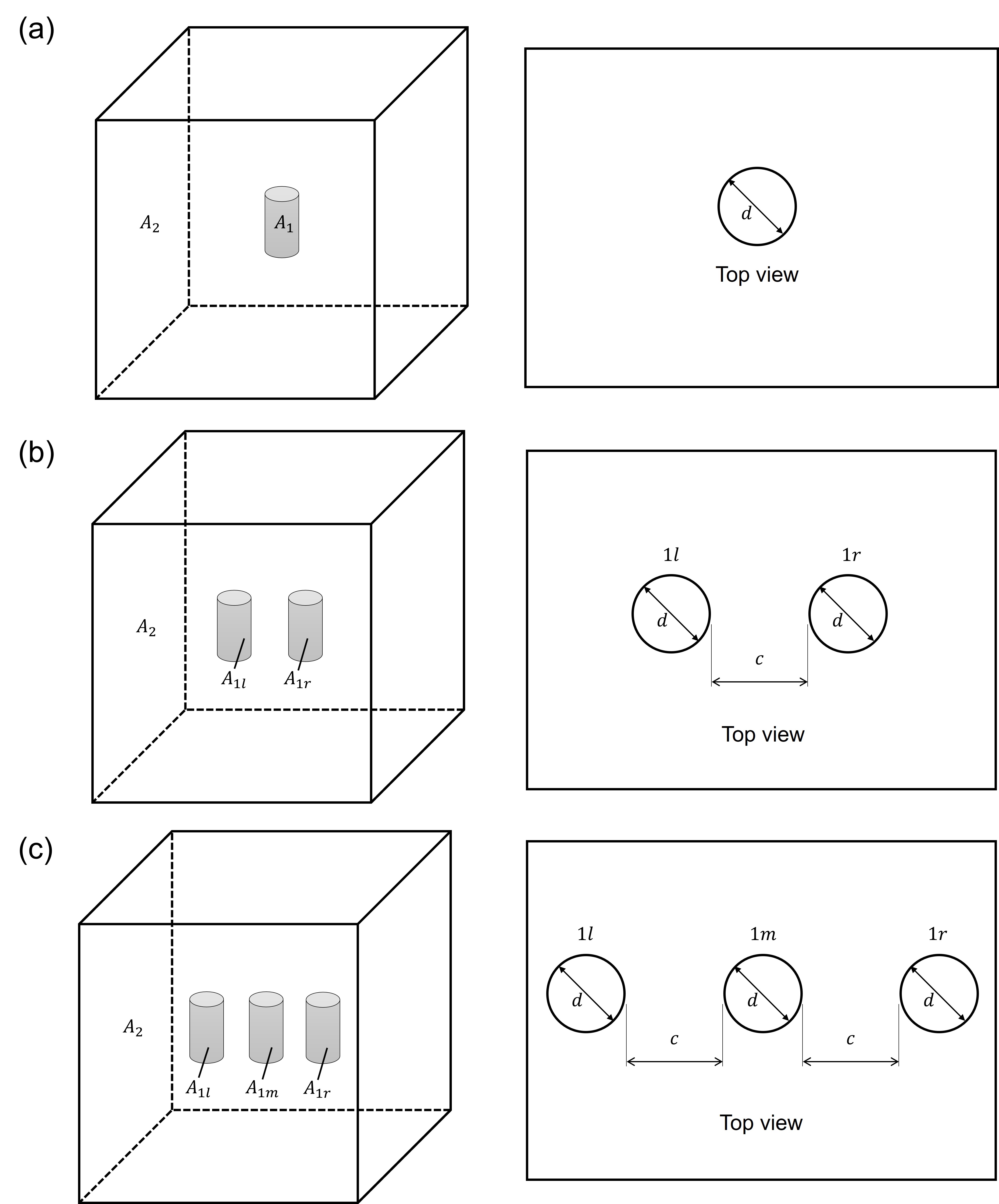}    
\caption{Three simple geometries where the view factor can be calculated analytically, including (a) a single vial, (b) two vials, and (c) three vials. The heating shelf and other equipment are omitted for clarity.}
    \label{fig:SimpleGeometry}
\end{figure}

\begin{figure}[ht!]
    \centering
\includegraphics[scale=0.5]{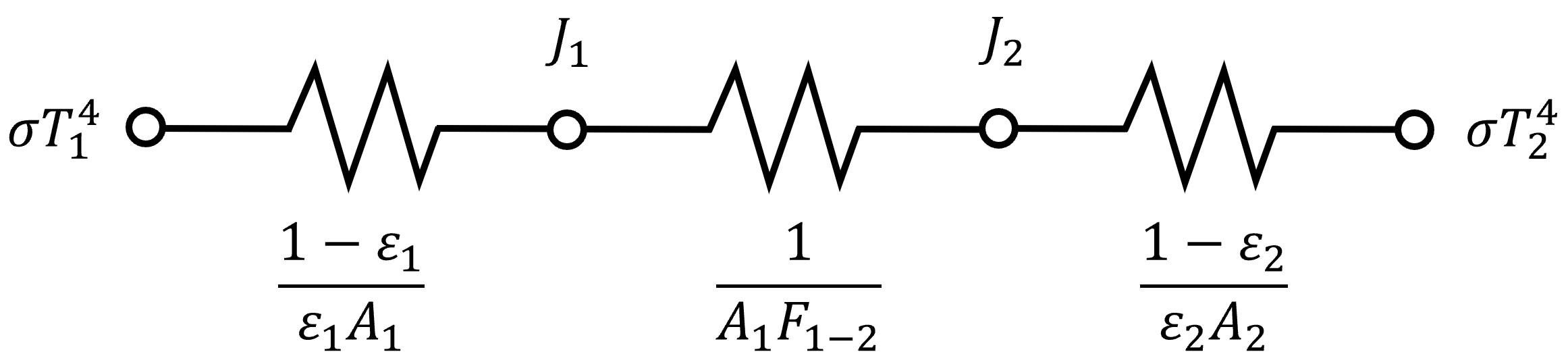}    
\caption{Equivalent electrical network for thermal radiation exchange between two surfaces.}
    \label{fig:NetworkDefault}
\end{figure}

\begin{figure}[ht!]
    \centering
\includegraphics[scale=.6]{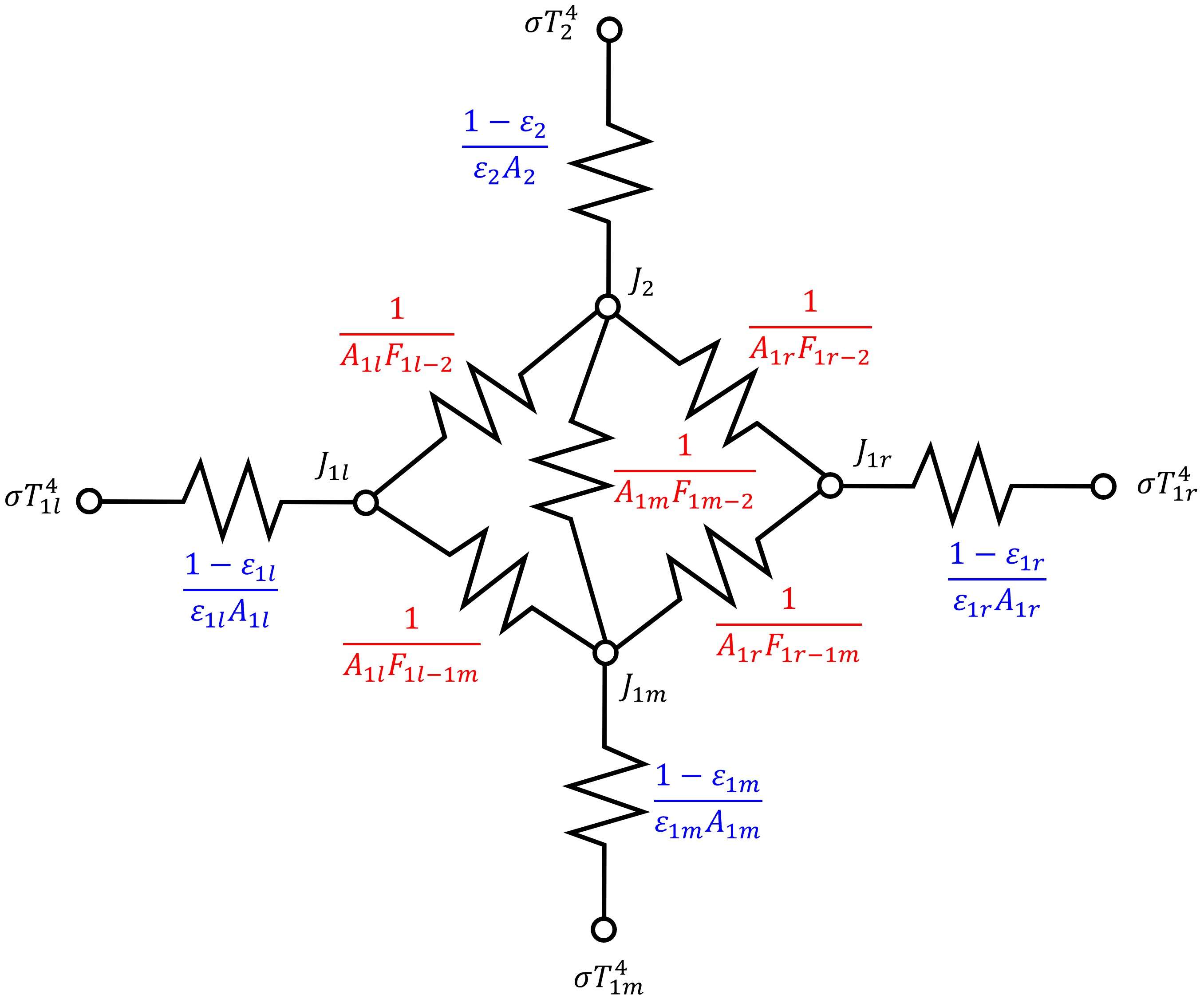}    
\caption{Network representation for radiation exchange between the four surfaces for the three-vial case in Fig.\ \ref{fig:SimpleGeometry}c. The surface resistances are highlighted in blue, while the space resistances are shown in red.}
    \label{fig:Network4Surfaces}
\end{figure}

\begin{figure}[ht!]
    \centering
\includegraphics[scale=.38]{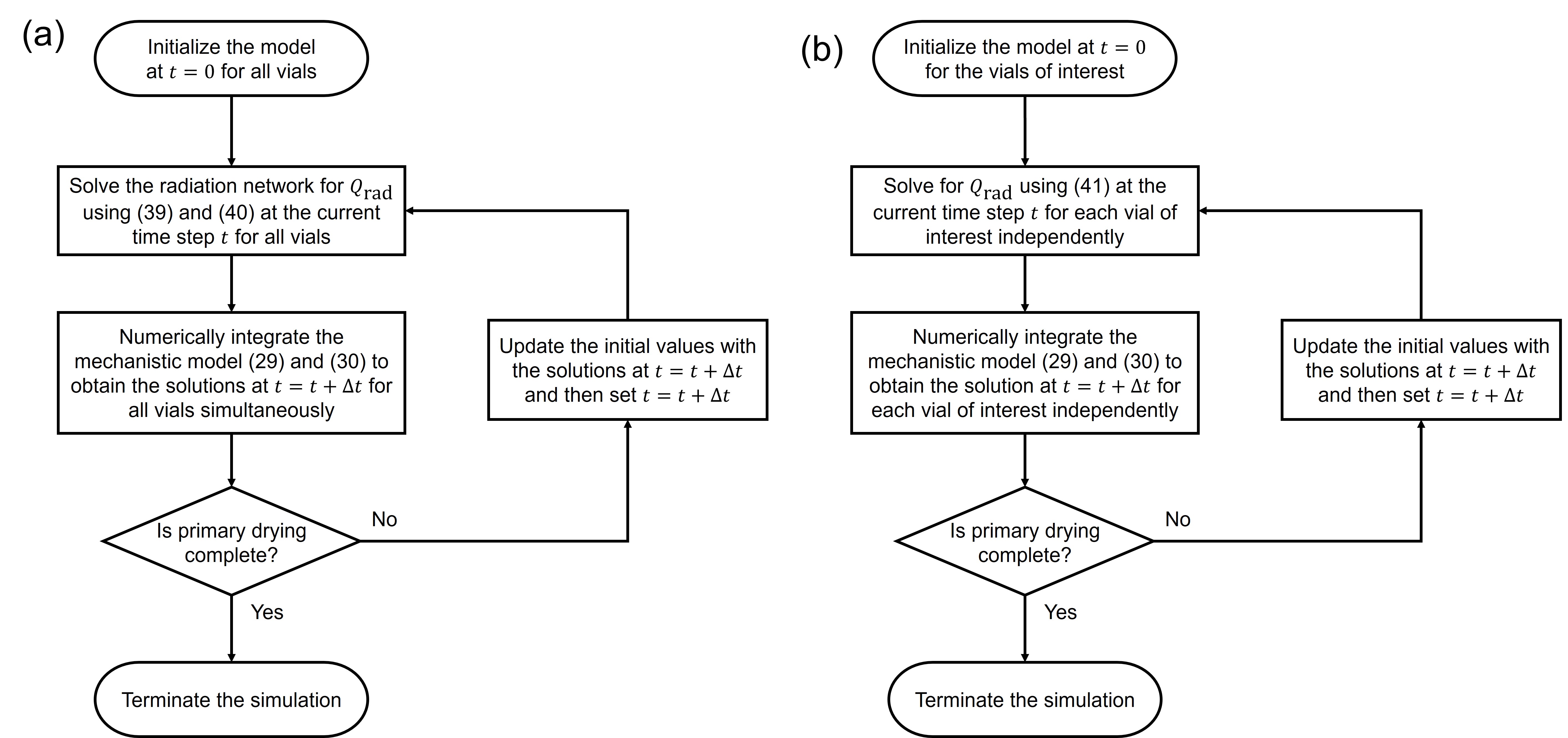}    
\caption{Flowcharts summarizing the calculation procedures for dynamic modeling of primary drying with the (a) radiation network and (b) simplified approaches.}
    \label{fig:Flowchart_combined}
\end{figure}

\begin{figure}[ht!]
    \centering
\includegraphics[scale=.72]{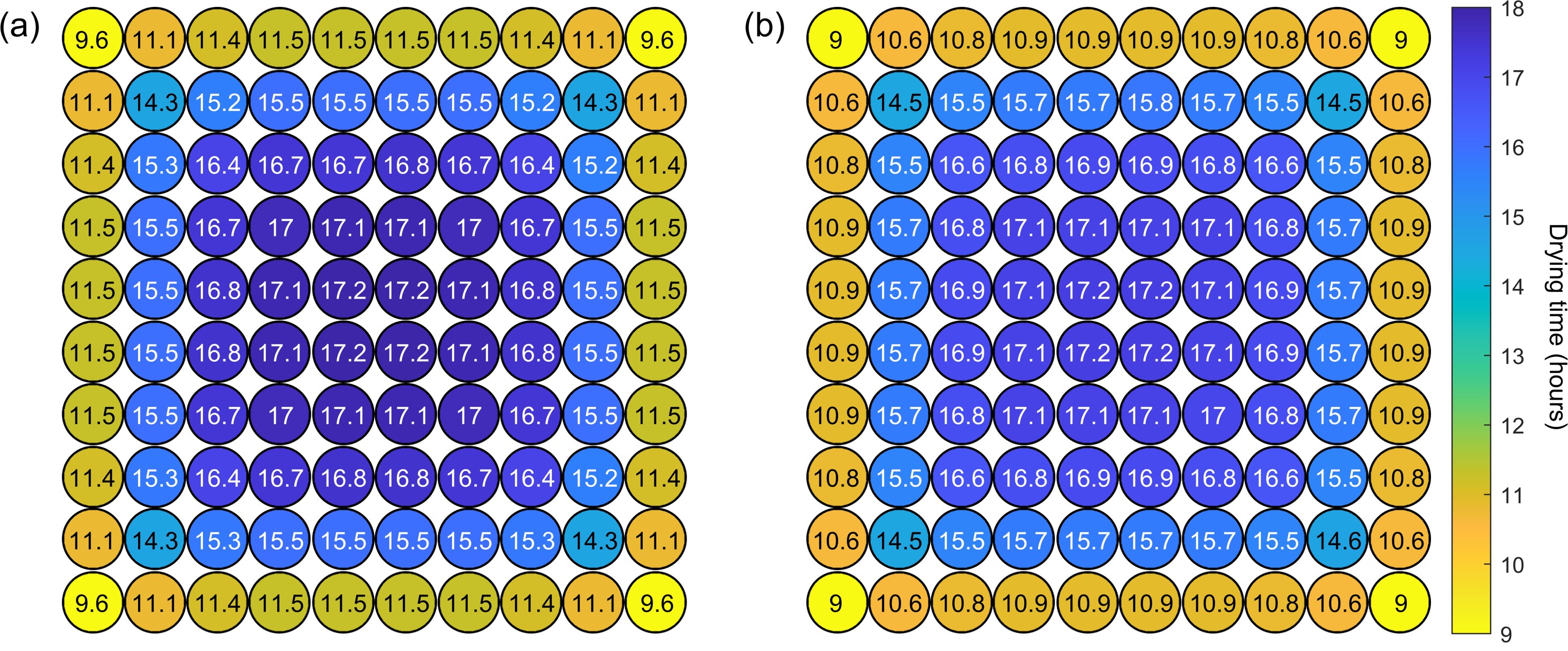}    
\caption{Comparison between the drying times predicted by the (a) radiation network and (b) simplified approaches for an array of 10$\times$10 vials. The drying time is 17.7 hours if thermal radiation is omitted.}
    \label{fig:Vial_10by10_Comparison}
\end{figure}

\begin{figure}[ht!]
    \centering
\includegraphics[scale=1]{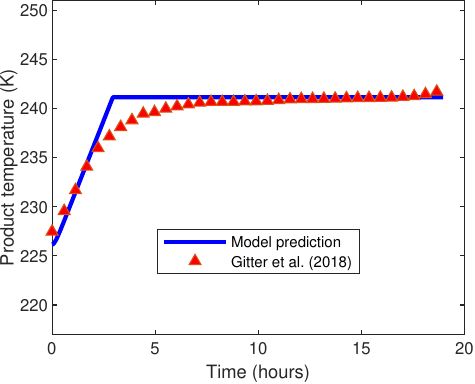}    
\caption{Comparison between the model prediction and experimental data for the product temperature at the bottom surface, Case 1.}
    \label{fig:Case1}
\end{figure}

\begin{figure}[ht!]
    \centering
\includegraphics[scale=.9]{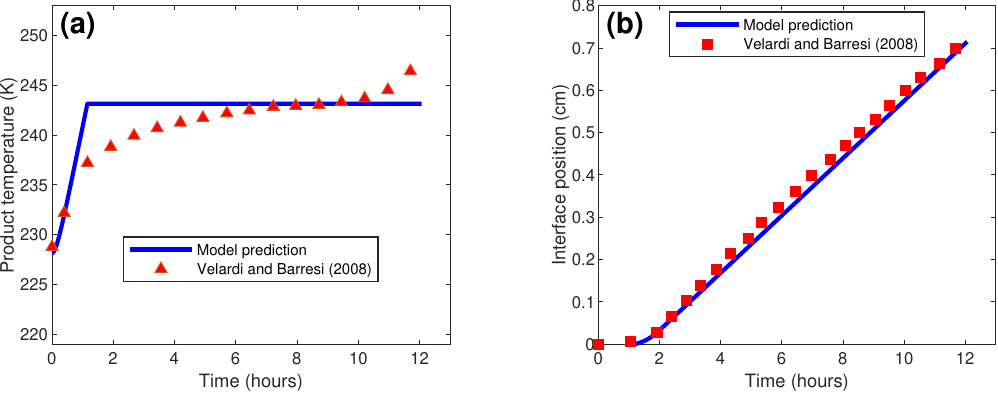}    
\caption{Comparison between the model prediction and (a) experimental data for the product temperature at the bottom surface and (b) the prediction of the model of Ref.\  \cite{Veraldi_2008_Parameters} for the interface position, Case 2 with the shelf temperature of 258.15 K. The drying time of $\sim$12 hours predicted by our model is somewhat larger than the experimental and model drying times reported by Ref.\ \cite{Veraldi_2008_Parameters}.}
    \label{fig:Case2_1}
\end{figure}

\begin{figure}[ht!]
    \centering
\includegraphics[scale=.9]{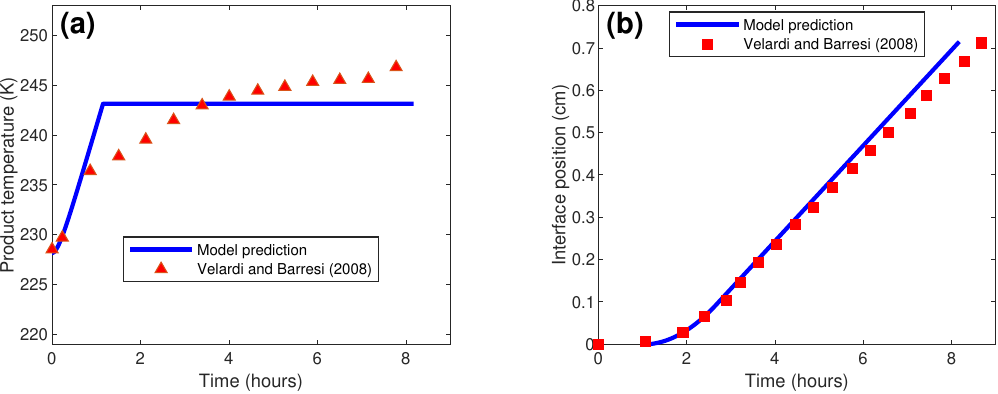}    
\caption{Comparison between the model prediction and (a) experimental data for the product temperature at the bottom surface and (b) the prediction of the model of Ref.\ \cite{Veraldi_2008_Parameters} for the interface position, Case 2 with the shelf temperature of 268.15 K. The drying time predicted by our model is somewhat larger than the experimental drying time, and somewhat smaller than the model drying time reported by Ref.\ \cite{Veraldi_2008_Parameters}.}.
    \label{fig:Case2_2}
\end{figure}

\begin{figure}[ht!]
    \centering
\includegraphics[scale=.9]{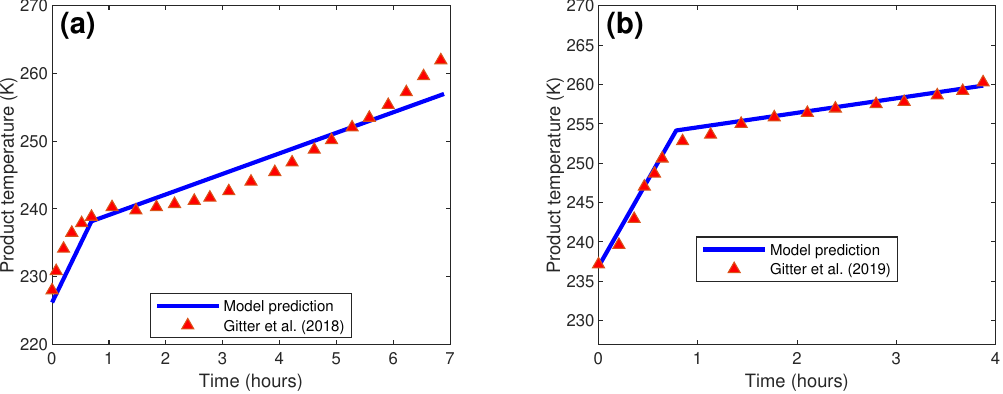}    
\caption{Comparison between the model prediction and experimental data for the product temperature at the bottom surface, (a) Case 3 and (b) Case 4.}
    \label{fig:Case3}
\end{figure}

\begin{figure}[ht!]
    \centering
\includegraphics[scale=0.67]{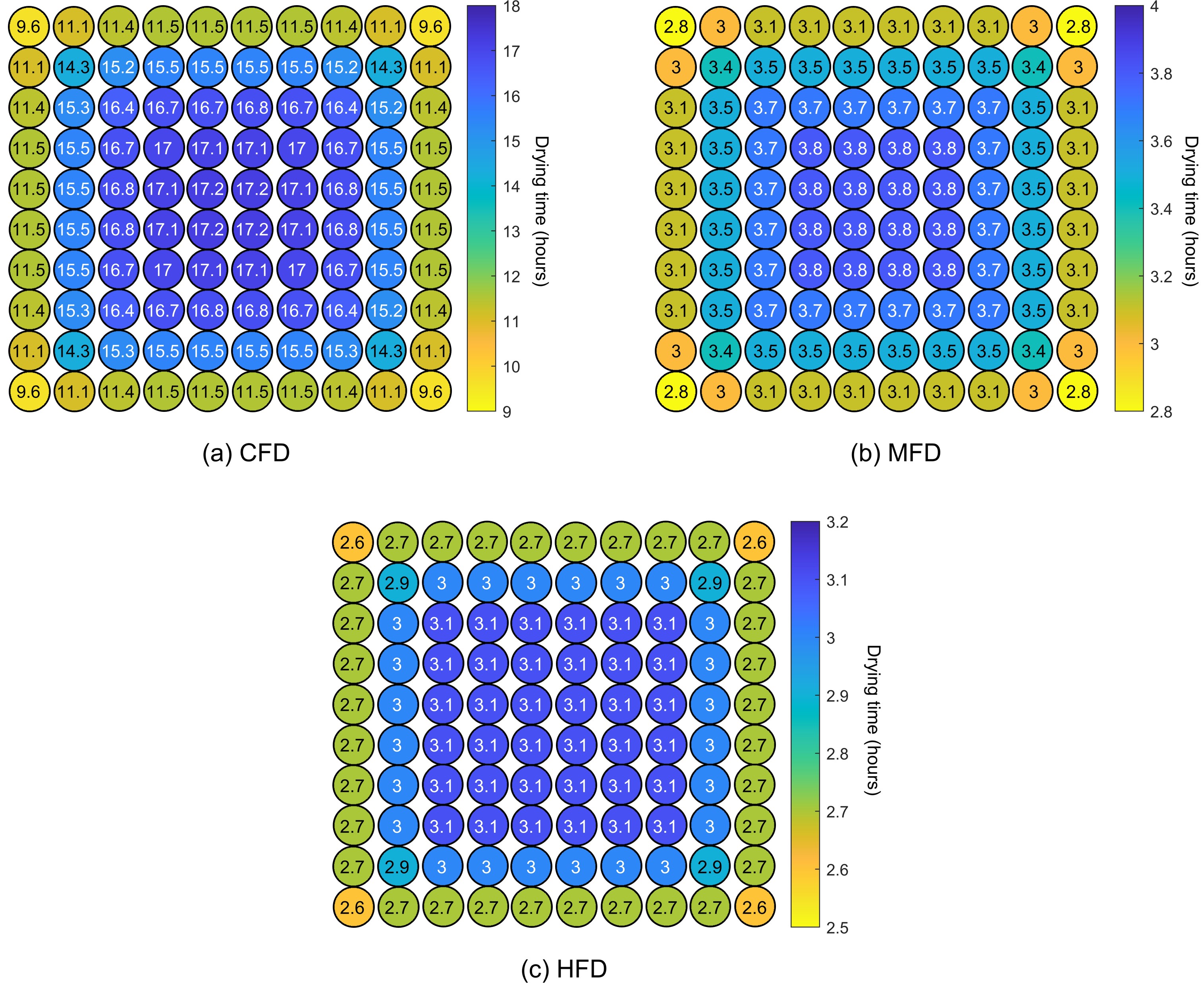}    
\caption{Drying times of an array of 10$\times$10 vials in (a) CFD, (b) MFD, and (c) HFD with thermal radiation. If thermal radiation is omitted, the drying times are 17.7, 4.0, and 3.2 hours for CFD, MFD, and HFD, respectively.}
    \label{fig:ThreeModes}
\end{figure}

\begin{figure}[ht!]
    \centering
\includegraphics[scale=0.9]{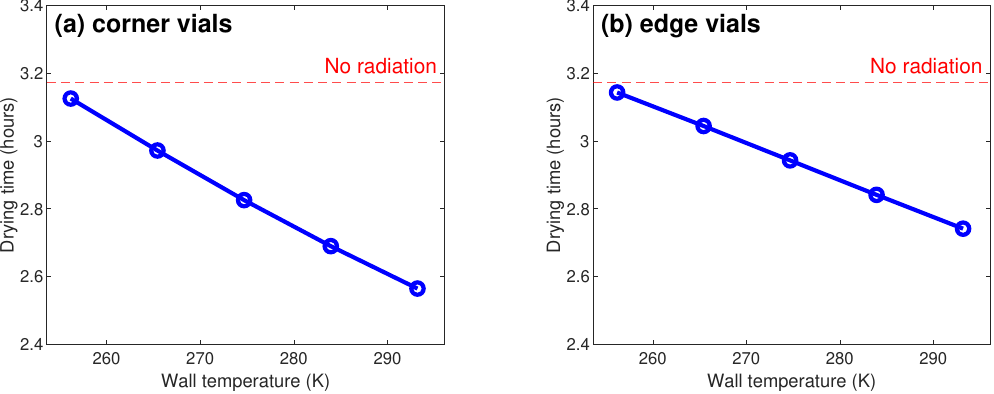}    
\caption{Effects of wall temperature on the drying times of the (a) corner vials and (b) edge vials in an array of 10$\times$10 vials in HFD.}
    \label{fig:WallTemp_HFD}
\end{figure}

\begin{figure}[ht!]
    \centering
\includegraphics[scale=0.75]{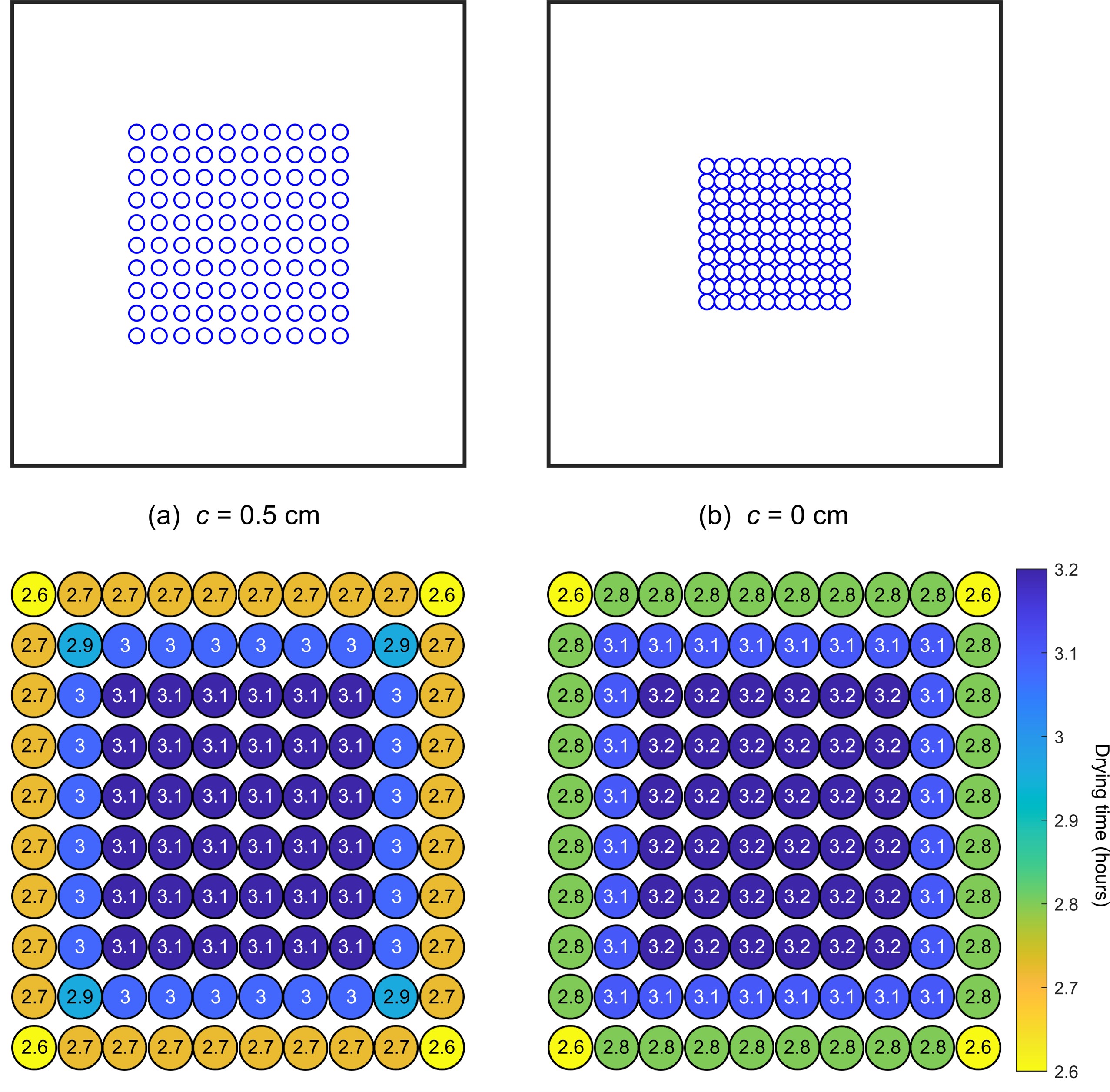}    
\caption{Drying times of an array of 10$\times$10 vials for (a) $c = 0.5$ cm and (b) $c=0$ cm in HFD with thermal radiation. The drying time is 3.2 hours if thermal radiation is omitted.}
    \label{fig:Dispo_distance}
\end{figure}

\begin{figure}[ht!]
    \centering
\includegraphics[scale=0.75]{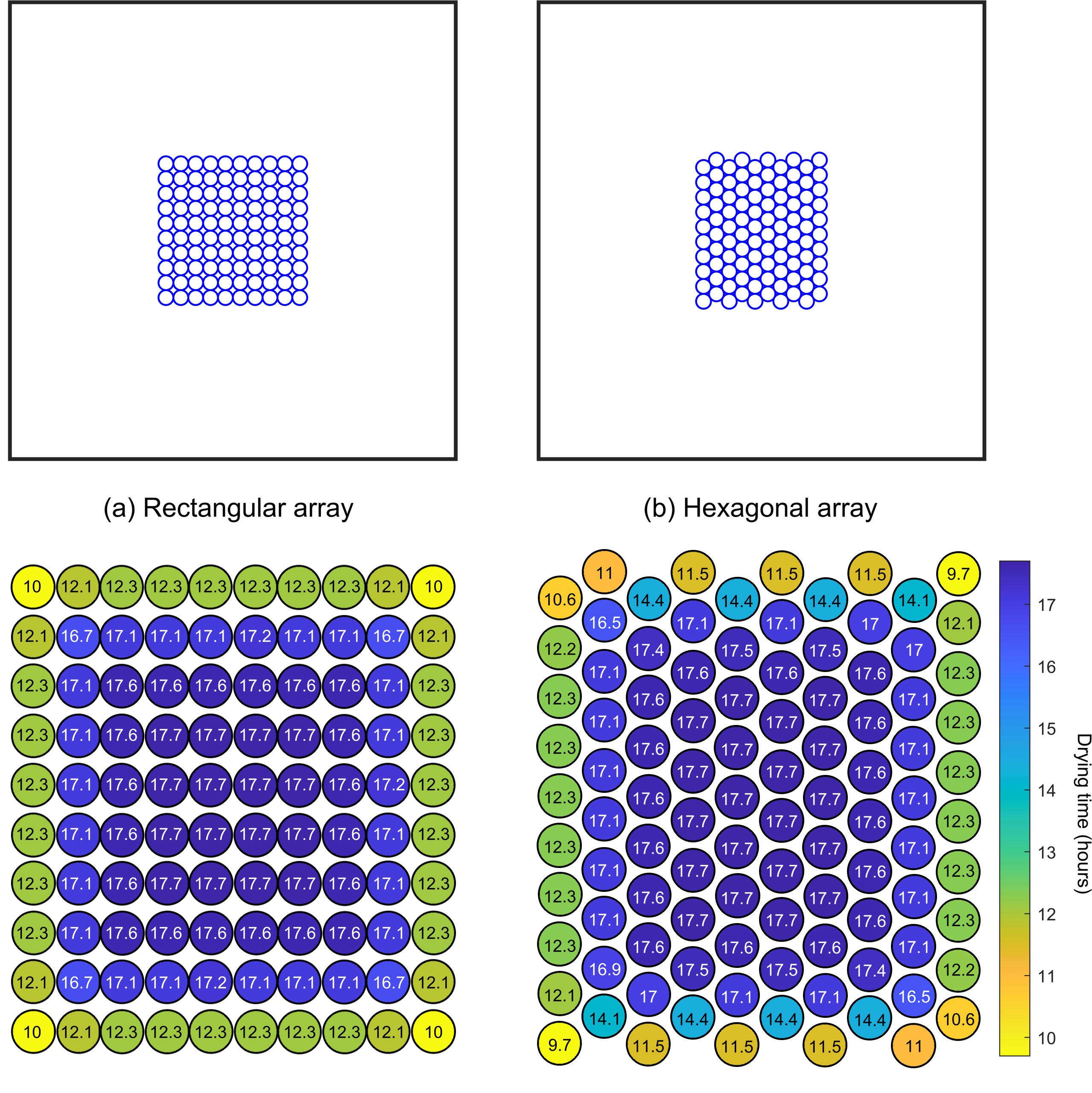}    
\caption{Drying times of the (a) rectangular array of 10$\times$10 vials and (b) hexagonal array of 10$\times$10 vials in CFD with thermal radiation, $c=0$ cm. The drying time is 17.7 hours if thermal radiation is omitted.}
    \label{fig:Dispo_CP}
\end{figure}

\begin{figure}[ht!]
    \centering
\includegraphics[scale=0.85]{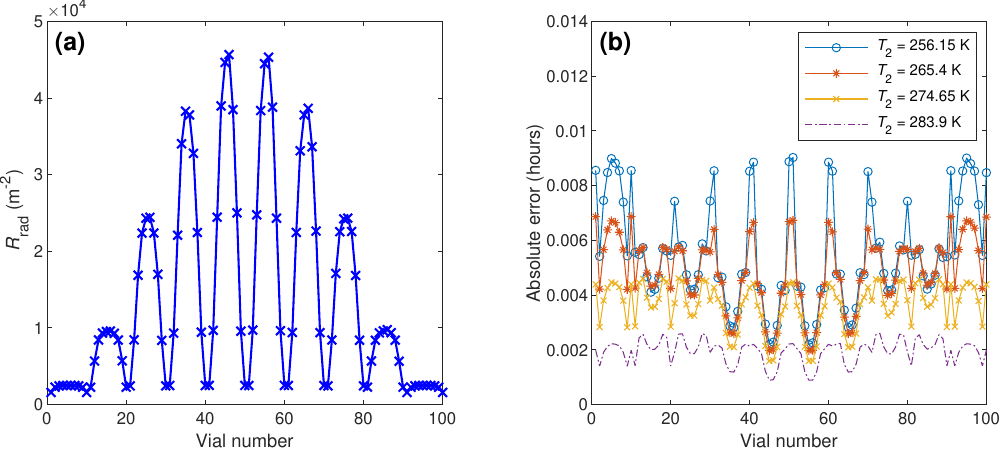}    
\caption{(a) Estimated $R_\textrm{rad}$ for a rectangular array of 10$\times$10 vials. (b) Errors between the drying times predicted by the radiation network (ground truth) and hybrid approaches. Vial numbering goes from the left to the right and from the bottom row to the top row, respectively; i.e., vial number 1 corresponds to the vial at the bottom left corner, while vial number 100 corresponds to the vial at the top right corner.}
    \label{fig:HybridModel}
\end{figure}

\clearpage

\begin{table}[h]
\caption{Default parameters for simulations.}
\renewcommand{\arraystretch}{1.2}
\label{Tab:Parameters}
\centering
\begin{tabular}{| c | c | c |c |}
\hline
\textbf{Parameter} & \textbf{Value} & \textbf{Unit} & \textbf{Reference/Note} \\ 
\hline
$\rho$  & 917  & kg/m$^3$ &   \cite{Veraldi_2008_Parameters}  \\ 
$\rho_d$ & 63 & kg/m$^3$ &   \cite{Veraldi_2008_Parameters}  \\
$k$ & 2.30 & W/m\mbox{-}K  &  \cite{Veraldi_2008_Parameters} \\  
$C_P$ & 1,967.8  & J/kg\mbox{-}K  &   \cite{Veraldi_2008_Parameters} \\ 
$\Delta H_\textrm{sub}$ & $2.84$$\times$$10^6$  &  J/kg &    \cite{Veraldi_2008_Parameters}\\
$Q$ & 85 & W &  \cite{Gitter_2019_Parameters} \\
$p_1$ & $3.73$$\times$$10^{-4}$ &  -- &   Estimated from \cite{Richard_2021_MainModel} \\
$p_2$ & $8.62$$\times$$10^{-3}$ &  -- &   Estimated from \cite{Richard_2021_MainModel}\\
$p_3$ & $2.5$$\times$$10^{-5}$ &  -- &  Estimated from \cite{Gitter_2019_Parameters}  \\
$h$ & 65 & W/m$^2$K &    \cite{Hottot_2006_Parameters} \\ 
$T_{0}$ & 236.85 & K &  \cite{Gitter_2019_Parameters}\\ 
$T_{b0}$ & 236.85 & K &   \cite{Gitter_2019_Parameters}\\ 
$T_{b,\textrm{max}}$ & 281.85 & K &  \cite{Richard_2021_MainModel} \\
$T_m$ & 256.15 &  K & \cite{Gitter_2019_Parameters} \\
$T_2$  & 293.15  & K & \cite{Gan_2005_WallTemp} \\ 
$r$ & 1 &  K/min &   \cite{Richard_2021_MainModel} \\
$L$ & 4.2 & cm &  \cite{Richard_2021_MainModel} \\ 
$d$ & 1 &  cm & --  \\
$c$ & 0.5 &  cm & -- \\
$A_1$ & 1.3$\times$10$^{-3}$ &  m$^2$ & Product of diameter 1 cm, height 4.2 cm \\
$A_2$ & 0.54 &  m$^2$ & Cubic of side 30 cm  \\
$V$ & $3.3$$\times$$10^{-6}$ & m$^3$ &  Product of diameter 1 cm, height 4.2 cm   \\
$\varepsilon_1$ & 0.8  & -- &  Glass, \cite{Mills_1995_HeatTransfer} \\ 
$\varepsilon_2$ & 0.3 $^a$  & -- & Stainless steel, \cite{Mills_1995_HeatTransfer,Pikal_2016_Convection} \\ 
$\sigma$ & $5.67$$\times$$10^{-8}$  & W/m$^2$K$^4$ & -- \\ 
\hline
\end{tabular}
\renewcommand{\arraystretch}{1}
\end{table}
{\footnotesize $^a$ Various values of the emissivity have been reported in the literature \cite{Pikal_2016_Convection}.}

\begin{table}[ht!]
\caption{Comparison between the view factors obtained from the analytical solutions and Monte Carlo method. The diameter of each vial ($d$) is 1 cm. The distance between vials ($c$) is 0.5 cm. The chamber is assumed to be a cube of side 30 cm.}
\renewcommand{\arraystretch}{1.2}
\label{Tab:MonteCarlo}
\centering
\begin{tabular}{|c|c|c|c|c|}
    \hline 
\textbf{Case}  & \textbf{View factor} &  \textbf{Analytical solution} & \textbf{Monte Carlo method} & \textbf{Error (\%)} \\ \hline
2 vials & $F_{1-2}$ & 0.8893 & 0.8883 & 0.11 \\ \hline 
\multirow{3}{*}{3 vials}  & $F_{1l-2}$ & 0.8893 & 0.8897 & 0.04 \\ \cline{2-5} 
    & $F_{1m-2}$ & 0.7786 & 0.7769 & 0.22 \\  \cline{2-5} 
    & $F_{1r-2}$ & 0.8893 & 0.8884 & 0.10 \\ \hline 
\end{tabular} 
\renewcommand{\arraystretch}{1}
\end{table}

\clearpage
\renewcommand{\arraystretch}{1.2}
\begin{longtable}[ht!]{| c | c | c | c | c |} 
\caption{Specific parameters for simulations in each case study.} 
\label{Tab:ParametersSpecific}\\
\hline
\textbf{Case study}&\textbf{Parameter} & \textbf{Value} & \textbf{Unit} & \textbf{Reference/Note} \\ 
\hline \endfirsthead
\multirow{9}{*}{Case 1} & $h$ & 19 & W/m$^2$K &  Estimated from data \\ 
& $T_{0}$ & 226.15 & K &   \cite{Gitter_2018_Experiment}\\ 
& $T_{b0}$ & 226.15 & K &   \cite{Gitter_2018_Experiment}\\ 
& $T_{b,\textrm{max}}$ & 253.15 & K &   \cite{Gitter_2018_Experiment} \\
& $T_m$ & 241.15 &  K &  \cite{Gitter_2018_Experiment} \\
& $r$ & 0.0889 & K/min &   \cite{Gitter_2018_Experiment}\\ 
& $L$ & 0.51 & cm &  Calculated from $V$ and $d$ \\ 
& $d$ & 2.4 & cm &   10R vial, \cite{Gitter_2018_Experiment} \\ 
& $V$  & 2.3  & mL &  \cite{Gitter_2018_Experiment} \\ \hline
\multirow{8}{*}{Case 2} & $\rho_d$ & 252 & kg/m$^3$ &  \cite{Veraldi_2008_Parameters} \\
& $h$ & 23.7 & W/m$^2$K &  Estimated from data \\
& $T_{0}$ & 228.15 & K &  \cite{Veraldi_2008_Parameters}  \\ 
& $T_{b0}$ & 228.15 & K &  \cite{Veraldi_2008_Parameters}\\ 
& $T_{b,\textrm{max}}$ & 258.15, 268.15 & K &   \cite{Veraldi_2008_Parameters} \\
& $T_m$ & 243.15 &  K &  Estimated from data \\
& $r$ & 0.25 & K/min &   \cite{Veraldi_2008_Parameters} \\ 
& $L$ & 0.715 & cm &  \cite{Veraldi_2008_Parameters} \\ \hline
\multirow{9}{*}{Case 3} &$Q$ & 25 &  W &  \cite{Gitter_2018_Experiment}  \\ 
& $p_1$ & $8.0$$\times$$10^{-4}$ &  -- &  Estimated from data  \\ 
& $p_2$ & $1.0$$\times$$10^{-2}$&  -- &  Estimated from data  \\ 
& $p_3$ & $1.4$$\times$$10^{-4}$&  -- &  Estimated from data  \\ 
& $T_{0}$ & 226.15 & K &   \cite{Gitter_2018_Experiment}\\ 
& $T_m$ & 238.15 &  K &  \cite{Gitter_2018_Experiment} \\
& $L$ &  0.51 & cm &  Calculated from $V$ and $d$  \\ 
& $d$ & 2.4 & cm &   10R vial, \cite{Gitter_2018_Experiment}  \\ 
& $V$  &  2.3 & mL &  \cite{Gitter_2018_Experiment}  \\ \hline
\multirow{9}{*}{Case 4} &$Q$ & 85 &  W &  \cite{Gitter_2019_Parameters}  \\ 
& $p_1$ & $3.0$$\times$$10^{-4}$ &  -- &  Estimated from data  \\ 
& $p_2$ &$5.9$$\times$$10^{-3}$&  -- &  Estimated from data  \\ 
& $p_3$ &  $2.5$$\times$$10^{-5}$ &  -- &  Estimated from data  \\ 
& $T_{0}$ & 236.85  & K & \cite{Gitter_2019_Parameters} \\ 
& $T_m$ & 254.15 &  K &  \cite{Gitter_2019_Parameters} \\
& $L$ &  0.51 & cm &  Calculated from $V$ and $d$  \\ 
& $d$ & 2.4 & cm &   10R vial, \cite{Gitter_2019_Parameters}  \\ 
& $V$  &  2.3 & mL & \cite{Gitter_2019_Parameters}   \\ \hline
\multirow{14}{*}{Case 5} & $\rho_d$ & 252 & kg/m$^3$ &  \cite{Veraldi_2008_Parameters} \\ 
&$h$ & 18.1 & W/m$^2$K &  Estimated from data \\
&$T_{0}$ & 230.75 & K &  \cite{Veraldi_2008_Parameters}  \\ 
&$T_{b0}$ & 230.75 & K &  \cite{Veraldi_2008_Parameters}\\ 
&$T_{b,\textrm{max}}$ & 268.15 & K &   \cite{Veraldi_2008_Parameters} \\
&$T_m$ & 242.70 &  K &  \cite{Veraldi_2008_Parameters} \\
&$T_2$ & 277.33 & K &  Estimated from data \\ 
&$r$ & 0.208 & K/min &   \cite{Veraldi_2008_Parameters} \\ 
&$L$ & 0.8 & cm &  \cite{Veraldi_2008_Parameters} \\ 
&$d$ & 1.425 & cm &  \cite{Veraldi_2008_Parameters} \\ 
&$c$ & 0 & cm &  \cite{Veraldi_2008_Parameters} \\ 
&$A_1$ & $3.58$$\times$$10^{-4}$ & m$^2$ &  Calculated from $L$ and $d$ \\ 
&$V$ & $1.28$$\times$$10^{-6}$ & m$^3$ &   Calculated from $L$ and $d$  \\ \hline
\multirow{11}{*}{Case 6} & $h$ & 24.8 & W/m$^2$K &  Estimated from data \\
&$T_{0}$ & 260 & K &  Assumed to be equal to $T_{b0}$  \\ 
&$T_{b0}$ & 260 & K &  \cite{Gan_2005_WallTemp} \\ 
&$T_{b,\textrm{max}}$ & 310 & K &  \cite{Gan_2005_WallTemp} \\
&$T_m$ & 263.15 &  K &  \cite{Gan_2005_WallTemp} \\
&$T_2$ & 293.15 & K & \cite{Gan_2005_WallTemp}  \\
&$L$ & 1.6 & cm & \cite{Gan_2005_WallTemp} \\ 
&$d$ & 1.4 & cm &  \cite{Gan_2005_WallTemp} \\ 
&$c$ & 0 & cm &  \cite{Gan_2005_WallTemp} \\ 
&$A_1$ & $7.04$$\times$$10^{-4}$ & m$^2$ &  Calculated from $L$ and $d$ \\
&$V$ & $2.46$$\times$$10^{-6}$ & m$^3$ &   Calculated from $L$ and $d$  \\ \hline
\end{longtable}
\renewcommand{\arraystretch}{1}

\begin{table}[ht!]
\caption{Comparison between the drying times predicted by our model and the reference data in \cite{Veraldi_2008_Parameters} for Case 5.}
\renewcommand{\arraystretch}{1.2}
\label{Tab:Result_Case4}
\centering
\begin{tabular}{|wc{7em}|wc{9em}|wc{9em}| wc{11em}|}
    \hline 
\multirow{2}{*}{\textbf{Case}}  &   \multicolumn{2}{c|}{\textbf{Total drying time (hours)}} & \multirow{2}{*}{\textbf{Note}}  \\ \cline{2-3} 
  &   \textbf{Reference data} & \textbf{Model prediction} &\\ \hline
No Radiation & 11.1 & 11.1 & --  \\ 
With Radiation  & 7.7 & 7.4 & Default $T_2 = 293.15$ K \\  
With Radiation  & 7.7 & 7.7 & Estimated $T_2 = 288.80$ K\\  \hline 
\end{tabular} 
\renewcommand{\arraystretch}{1}
\end{table}

\begin{table}
\caption{Comparison between the drying times predicted by our model and the reference data in \cite{Gan_2005_WallTemp} for Case 6.}
\renewcommand{\arraystretch}{1.2}
\label{Tab:Result_Case5}
\centering
\begin{tabular}{|wc{7em}|wc{9em}|wc{9em}|}
    \hline 
\multirow{2}{*}{\textbf{Case}}  &   \multicolumn{2}{c|}{\textbf{Total drying time (hours)}}  \\ \cline{2-3} 
  &   \textbf{Reference data} & \textbf{Model prediction} \\ \hline
Center vial & 9.74 & 9.74  \\ 
Edge vial  & 8.37 & 8.47 \\  
Corner vial  & 7.89 & 7.61 \\  \hline 
\end{tabular} 
\renewcommand{\arraystretch}{1}
\end{table}

\begin{table}
\caption{Comparison of the total radiative heat transfer in different vials and freeze-drying methods.}
\renewcommand{\arraystretch}{1.2}
\label{Tab:Heat_rad}
\centering
\begin{tabular}{|wc{6em}|wc{7em}|wc{7em}|wc{7em}|}
    \hline 
\multirow{2}{*}{\textbf{Method}}  &   \multicolumn{3}{c|}{\textbf{Total radiative heat transfer in the vials (J)}}  \\ \cline{2-4} 
  &   \textbf{Corner vials} & \textbf{Edge vials} & \textbf{Center vials} \\ \hline
CFD & 3,975 & 3,014 & 184 \\ 
MFD  & 1,154 & 782 & 63\\   
HFD  & 1,073 & 725 & 41 \\  \hline 
\end{tabular} 
\renewcommand{\arraystretch}{1}
\end{table}

\begin{table}
\caption{Drying times of the corner and edge vials for six different vial layouts. All vials are arranged in a rectangular array.}
\renewcommand{\arraystretch}{1.2}
\label{Tab:Dispo_layout}
\centering
\begin{tabular}{|wc{7em}|wc{8em}|wc{8em}|}
    \hline 
\multirow{2}{*}{\textbf{Layout}}  &   \multicolumn{2}{c|}{\textbf{Total drying time (hours)}}  \\ \cline{2-3} 
  &   \textbf{Corner vials} & \textbf{Edge vials} \\ \hline
1 vial & 2.34 & --  \\ 
2$\times$2 vials  & 2.48 & -- \\  
5$\times$5 vials  & 2.54 & 2.68 \\  
8$\times$8 vials  & 2.56 & 2.73 \\  
10$\times$10 vials   & 2.56 & 2.74 \\  
15$\times$15 vials  & 2.59 & 2.76 \\  \hline 
\end{tabular} 
\renewcommand{\arraystretch}{1}
\end{table}

\clearpage

\clearpage
\bibliography{reference}

\begin{thebibliography}{10}
\expandafter\ifx\csname url\endcsname\relax
  \def\url#1{\texttt{#1}}\fi
\expandafter\ifx\csname urlprefix\endcsname\relax\def\urlprefix{URL }\fi
\expandafter\ifx\csname href\endcsname\relax
  \def\href#1#2{#2} \def\path#1{#1}\fi

\bibitem{Fissore_2018_Review}
D.~Fissore, R.~Pisano, A.~A. Barresi, Process analytical technology for monitoring pharmaceuticals freeze-drying – a comprehensive review, Dry. Technol. 36~(15) (2018) 1839--1865.
\newblock \href {https://doi.org/10.1080/07373937.2018.1440590} {\path{doi:10.1080/07373937.2018.1440590}}.

\bibitem{Bano_2020_optimize}
G.~Bano, R.~De-Luca, E.~Tomba, A.~Marcelli, F.~Bezzo, M.~Barolo, Primary drying optimization in pharmaceutical freeze-drying: A multivial stochastic modeling framework, Ind. Eng. Chem. Res. 59~(11) (2020) 5056--5071.
\newblock \href {https://doi.org/10.1021/acs.iecr.9b06402} {\path{doi:10.1021/acs.iecr.9b06402}}.

\bibitem{Pisano_2010_control}
R.~Pisano, D.~Fissore, S.~A. Velardi, A.~A. Barresi, In-line optimization and control of an industrial freeze-drying process for pharmaceuticals, J. Pharm. Sci. 99~(11) (2010) 4691--4709.
\newblock \href {https://doi.org/10.1002/jps.22166} {\path{doi:10.1002/jps.22166}}.

\bibitem{Veraldi_2008_Parameters}
S.~A. Velardi, A.~A. Barresi, Development of simplified models for the freeze-drying process and investigation of the optimal operating conditions, Chem. Eng. Res. Des. 86~(1) (2008) 9--22.
\newblock \href {https://doi.org/10.1016/j.cherd.2007.10.007} {\path{doi:10.1016/j.cherd.2007.10.007}}.

\bibitem{Barresi_2009_Monitoring}
A.~A. Barresi, R.~Pisano, D.~Fissore, V.~Rasetto, S.~A. Velardi, A.~Vallan, M.~Parvis, M.~Galan, Monitoring of the primary drying of a lyophilization process in vials, Chem. Eng. Process. 48~(1) (2009) 408--423.
\newblock \href {https://doi.org/10.1016/j.cep.2008.05.004} {\path{doi:10.1016/j.cep.2008.05.004}}.

\bibitem{Muramatsu_2022_mRNA}
H.~Muramatsu, K.~Lam, C.~Bajusz, D.~Laczkó, K.~Karikó, P.~Schreiner, A.~Martin, P.~Lutwyche, J.~Heyes, N.~Pardi, Lyophilization provides long-term stability for a lipid nanoparticle-formulated, nucleoside-modified m{RNA} vaccine, Mol. Ther. 30~(5) (2022) 1941--1951.
\newblock \href {https://doi.org/10.1016/j.ymthe.2022.02.001} {\path{doi:10.1016/j.ymthe.2022.02.001}}.

\bibitem{Meulewaeter_2023_mRNA}
S.~Meulewaeter, G.~NuFytten, M.~H. Cheng, S.~C. {De Smedt}, P.~R. Cullis, T.~{De Beer}, I.~Lentacker, R.~Verbeke, Continuous freeze-drying of messenger {RNA} lipid nanoparticles enables storage at higher temperatures, J. Control. Release 357 (2023) 149--160.
\newblock \href {https://doi.org/10.1016/j.jconrel.2023.03.039} {\path{doi:10.1016/j.jconrel.2023.03.039}}.

\bibitem{Gitter_2018_Experiment}
J.~H. Gitter, R.~Geidobler, I.~Presser, G.~Winter, Significant drying time reduction using microwave-assisted freeze-drying for a monoclonal antibody, J. Pharm. Sci. 107~(10) (2018) 2538--2543.
\newblock \href {https://doi.org/10.1016/j.xphs.2018.05.023} {\path{doi:10.1016/j.xphs.2018.05.023}}.

\bibitem{Gitter_2019_Parameters}
J.~H. Gitter, R.~Geidobler, I.~Presser, G.~Winter, Microwave-assisted freeze-drying of monoclonal antibodies: Product quality aspects and storage stability, Pharmaceutics 11~(12) (2019) 674.
\newblock \href {https://doi.org/10.3390/pharmaceutics11120674} {\path{doi:10.3390/pharmaceutics11120674}}.

\bibitem{Bhambhani_2021_MVD}
A.~Bhambhani, J.~Stanbro, D.~Roth, E.~Sullivan, M.~Jones, R.~Evans, J.~Blue, Evaluation of microwave vacuum drying as an alternative to freeze-drying of biologics and vaccines: The power of simple modeling to identify a mechanism for faster drying times achieved with microwave, AAPS PharmSciTech 22~(1) (2021) 52.
\newblock \href {https://doi.org/10.1208/s12249-020-01912-9} {\path{doi:10.1208/s12249-020-01912-9}}.

\bibitem{Richard_2021_MainModel}
J.~Park, J.~H. Cho, R.~D. Braatz, Mathematical modeling and analysis of microwave-assisted freeze-drying in biopharmaceutical applications, Comput. Chem. Eng. 153 (2021) 107412.
\newblock \href {https://doi.org/10.1016/j.compchemeng.2021.107412} {\path{doi:10.1016/j.compchemeng.2021.107412}}.

\bibitem{Litchfield_1979_Model}
R.~Litchfield, A.~Liapis, An adsorption-sublimation model for a freeze dryer, Chem. Eng. Sci. 34~(9) (1979) 1085--1090.
\newblock \href {https://doi.org/10.1016/0009-2509(79)85013-7} {\path{doi:10.1016/0009-2509(79)85013-7}}.

\bibitem{Mascarenhas_1997_FEMmodel}
W.~Mascarenhas, H.~Akay, M.~Pikal, A computational model for finite element analysis of the freeze-drying process, Comput. Methods Appl. Mech. Eng. 148~(1) (1997) 105--124.
\newblock \href {https://doi.org/10.1016/S0045-7825(96)00078-3} {\path{doi:10.1016/S0045-7825(96)00078-3}}.

\bibitem{Pikal_2005_Model}
M.~J. Pikal, W.~J. Mascarenhas, H.~U. Akay, S.~Cardon, C.~Bhugra, F.~Jameel, S.~Rambhatla, The nonsteady state modeling of freeze drying: In-process product temperature and moisture content mapping and pharmaceutical product quality applications, Pharm. Dev. Technol. 10~(1) (2005) 17--32.
\newblock \href {https://doi.org/10.1081/PDT-35869} {\path{doi:10.1081/PDT-35869}}.

\bibitem{Hottot_2006_Parameters}
A.~Hottot, R.~Peczalski, S.~Vessot, J.~Andrieu, Freeze-drying of pharmaceutical proteins in vials: Modeling of freezing and sublimation steps, Dry. Technol. 24~(5) (2006) 561--570.
\newblock \href {https://doi.org/10.1080/07373930600626388} {\path{doi:10.1080/07373930600626388}}.

\bibitem{Nastaj_2009_MFD}
J.~Nastaj, K.~Witkiewicz, Mathematical modeling of the primary and secondary vacuum freeze drying of random solids at microwave heating, Int. J. Heat Mass Transf. 52~(21) (2009) 4796--4806.
\newblock \href {https://doi.org/10.1016/j.ijheatmasstransfer.2009.06.015} {\path{doi:10.1016/j.ijheatmasstransfer.2009.06.015}}.

\bibitem{Chen_2015_FEMmodel}
X.~Chen, V.~Sadineni, M.~Maity, Y.~Quan, M.~Enterline, R.~V. Mantri, Finite element method {(FEM)} modeling of freeze-drying: Monitoring pharmaceutical product robustness during lyophilization, AAPS PharmSciTech 16 (2015) 1317--1326.
\newblock \href {https://doi.org/10.1208/s12249-015-0318-9} {\path{doi:10.1208/s12249-015-0318-9}}.

\bibitem{Fissore_2015_HTCestimation}
D.~Fissore, R.~Pisano, A.~A. Barresi, Using mathematical modeling and prior knowledge for {Q}b{D} in freeze-drying processes, in: F.~Jameel, S.~Hershenson, M.~A. Khan, S.~Martin-Moe (Eds.), Quality by Design for Biopharmaceutical Drug Product Development, Springer, New York, 2015, pp. 565--593.
\newblock \href {https://doi.org/10.1007/978-1-4939-2316-8_23} {\path{doi:10.1007/978-1-4939-2316-8_23}}.

\bibitem{Scutella_2017_3Dmodel}
B.~Scutellà, A.~Plana-Fattori, S.~Passot, E.~Bourlès, F.~Fonseca, D.~Flick, I.~Tréléa, {3D} mathematical modelling to understand atypical heat transfer observed in vial freeze-drying, Appl. Therm. Eng. 126 (2017) 226--236.
\newblock \href {https://doi.org/10.1016/j.applthermaleng.2017.07.096} {\path{doi:10.1016/j.applthermaleng.2017.07.096}}.

\bibitem{Wang_2020_MircowaveModel}
W.~Wang, S.~Zhang, Y.~Pan, J.~Yang, Y.~Tang, G.~Chen, Multiphysics modeling for microwave freeze-drying of initially porous frozen material assisted by wave-absorptive medium, Ind. Eng. Chem. Res. 59~(47) (2020) 20903--20915.
\newblock \href {https://doi.org/10.1021/acs.iecr.0c03852} {\path{doi:10.1021/acs.iecr.0c03852}}.

\bibitem{Srisuma_2023_AnalyticalLyo}
P.~Srisuma, G.~Barbastathis, R.~D. Braatz, Analytical solutions for the modeling, optimization, and control of microwave-assisted freeze drying, Comput. Chem. Eng. 177 (2023) 108318.
\newblock \href {https://doi.org/10.1016/j.compchemeng.2023.108318} {\path{doi:10.1016/j.compchemeng.2023.108318}}.

\bibitem{Pisano_2011_HeatTransferInLyo}
R.~Pisano, D.~Fissore, A.~A. Barresi, Heat transfer in freeze-drying apparatus, in: M.~A. dos Santos~Bernardes (Ed.), Developments in Heat Transfer, IntechOpen, Rijeka, 2011, pp. 92--114.
\newblock \href {https://doi.org/10.5772/23799} {\path{doi:10.5772/23799}}.

\bibitem{Pikal_2016_Convection}
M.~J. Pikal, R.~Bogner, V.~Mudhivarthi, P.~Sharma, P.~Sane, Freeze-drying process development and scale-up: Scale-up of edge vial versus center vial heat transfer coefficients, ${K}_v$, J. Pharm. Sci. 105~(11) (2016) 3333--3343.
\newblock \href {https://doi.org/10.1016/j.xphs.2016.07.027} {\path{doi:10.1016/j.xphs.2016.07.027}}.

\bibitem{Rambhatla_2003_RadiationShield}
S.~Rambhatla, M.~J. Pikal, Heat and mass transfer scale-up issues during freeze-drying, {I}: Atypical radiation and the edge vial effect, AAPS PharmSciTech 4~(2) (2003) 14.
\newblock \href {https://doi.org/10.1208/pt040214} {\path{doi:10.1208/pt040214}}.

\bibitem{Gan_2005_WallTemp}
K.~Gan, R.~Bruttini, O.~Crosser, A.~Liapis, Freeze-drying of pharmaceuticals in vials on trays: Effects of drying chamber wall temperature and tray side on lyophilization performance, Int. J. Heat Mass Transf. 48~(9) (2005) 1675--1687.
\newblock \href {https://doi.org/10.1016/j.ijheatmasstransfer.2004.12.004} {\path{doi:10.1016/j.ijheatmasstransfer.2004.12.004}}.

\bibitem{Ehlers_2021_WallTemp}
S.~Ehlers, W.~Friess, R.~Schroeder, Impact of chamber wall temperature on energy transfer during freeze-drying, Int. J. Pharm. 592 (2021) 120025.
\newblock \href {https://doi.org/10.1016/j.ijpharm.2020.120025} {\path{doi:10.1016/j.ijpharm.2020.120025}}.

\bibitem{Sheehan_1998_Modeling}
P.~Sheehan, A.~I. Liapis, Modeling of the primary and secondary drying stages of the freeze drying of pharmaceutical products in vials: {N}umerical results obtained from the solution of a dynamic and spatially multi-dimensional lyophilization model for different operational policies, Biotechnol. Bioeng. 60~(6) (1998) 712--728.
\newblock \href {https://doi.org/10.1002/(SICI)1097-0290(19981220)60:6<712::AID-BIT8>3.0.CO;2-4} {\path{doi:10.1002/(SICI)1097-0290(19981220)60:6<712::AID-BIT8>3.0.CO;2-4}}.

\bibitem{Mills_1995_HeatTransfer}
A.~Mills, Heat and Mass Transfer, Routledge, New York, 1995.
\newblock \href {https://doi.org/10.4324/9780203752173} {\path{doi:10.4324/9780203752173}}.

\bibitem{Dryer_1969_HeatTransferControlled}
D.~Dryer, J.~Sunderland, The role of convection in drying, Chem. Eng. Sci. 23~(9) (1968) 965--970.
\newblock \href {https://doi.org/10.1016/0009-2509(68)87082-4} {\path{doi:10.1016/0009-2509(68)87082-4}}.

\bibitem{Jafar_2003_HeatTransferLimit}
F.~Jafar, M.~Farid, Analysis of heat and mass transfer in freeze drying, Dry. Technol. 21~(2) (2003) 249--263.
\newblock \href {https://doi.org/10.1081/DRT-120017746} {\path{doi:10.1081/DRT-120017746}}.

\bibitem{Pikal_1984_HTCexperiment}
M.~J. Pikal, M.~L. Roy, S.~Shah, Mass and heat transfer in vial freeze-drying of pharmaceuticals: Role of the vial, J. Pharm. Sci. 73~(9) (1984) 1224--1237.
\newblock \href {https://doi.org/10.1002/jps.2600730910} {\path{doi:10.1002/jps.2600730910}}.

\bibitem{Incropera_2007_HeatTransfer}
F.~P. Incropera, D.~P. Dewitt, T.~L. Bergman, A.~S. Lavine, Fundamentals of Heat and Mass Transfer, sixth Edition, John Wiley \& Sons, New Jersey, 2007.

\bibitem{Jiang_2020_NumerViewFactor}
C.~Jiang, J.~Wang, O.~Behar, C.~Caliot, Y.~Zhang, G.~Flamant, A modified numerical integration method to calculate the view factor between finite and infinite cylinders in arbitrary array, Ann. Nucl. Energy 142 (2020) 107358.
\newblock \href {https://doi.org/10.1016/j.anucene.2020.107358} {\path{doi:10.1016/j.anucene.2020.107358}}.

\bibitem{Mirhosseini_2011_MonteCarlo}
M.~Mirhosseini, A.~Saboonchi, View factor calculation using the {Monte Carlo} method for a {3D} strip element to circular cylinder, Int. Commun. Heat Mass Transf. 38~(6) (2011) 821--826.
\newblock \href {https://doi.org/10.1016/j.icheatmasstransfer.2011.03.022} {\path{doi:10.1016/j.icheatmasstransfer.2011.03.022}}.

\bibitem{Schwarz_2023_intersect}
D.~Schwarz, Fast and robust curve intersections, \url{https://www.mathworks.com/matlabcentral/fileexchange/11837-fast-and-robust-curve-intersections}, retrieved June 1, 2023 (2023).

\bibitem{Juul_1982_FiniteCylinder}
N.~H. Juul, View factors in radiation between two parallel oriented cylinders of finite lengths, J. Heat Transfer 104~(2) (1982) 384--388.
\newblock \href {https://doi.org/10.1115/1.3245101} {\path{doi:10.1115/1.3245101}}.

\bibitem{Abdelraheem_2022_Uniformity}
A.~Abdelraheem, R.~Tukra, P.~Kazarin, M.~D. Sinanis, E.~M. Topp, A.~Alexeenko, D.~Peroulis, {Statistical electromagnetics for industrial pharmaceutical lyophilization}, PNAS Nexus 1~(3) (2022) pgac052.
\newblock \href {https://doi.org/10.1093/pnasnexus/pgac052} {\path{doi:10.1093/pnasnexus/pgac052}}.

\bibitem{Hottot_2005_HTC}
A.~Hottot, S.~Vessot, J.~Andrieu, Determination of mass and heat transfer parameters during freeze-drying cycles of pharmaceutical products., PDA J. Pharm. Sci. Technol. 59~(2) (2005) 138--153.

\bibitem{Capozzi_2019_conlyo}
L.~C. Capozzi, B.~L. Trout, R.~Pisano, From batch to continuous: Freeze-drying of suspended vials for pharmaceuticals in unit-doses, Ind. Eng. Chem. Res. 58~(4) (2019) 1635--1649.
\newblock \href {https://doi.org/10.1021/acs.iecr.8b02886} {\path{doi:10.1021/acs.iecr.8b02886}}.

\bibitem{Pisano_2019_conlyo}
R.~Pisano, A.~Arsiccio, L.~C. Capozzi, B.~L. Trout, Achieving continuous manufacturing in lyophilization: Technologies and approaches, Eur. J. Pharm. Biopharm. 142 (2019) 265--279.
\newblock \href {https://doi.org/10.1016/j.ejpb.2019.06.027} {\path{doi:10.1016/j.ejpb.2019.06.027}}.

\end{thebibliography}
\end{document}